\documentclass[a4paper]{article}

\usepackage[parfill]{parskip} 
\usepackage[margin=1in]{geometry}
\usepackage[doublespacing]{setspace}
\usepackage[sort,round]{natbib}

\usepackage[pagebackref=true,bookmarks]{hyperref}
\hypersetup{
	unicode=false,
	pdftoolbar=true,
	pdfmenubar=true,
	pdffitwindow=false,     
	pdfstartview={FitH},    
	pdftitle={Penalized GLMM},    
	pdfauthor={JSP},     
	pdfsubject={Subject},   
	pdfcreator={JSP},   
	pdfproducer={JSP}, 
	pdfkeywords={}, 
	pdfnewwindow=true,      
	colorlinks=true,       
	linkcolor=red,          
	citecolor=blue,        
	filecolor=black,      
	urlcolor=blue           
}

\usepackage{url}
\usepackage{amsfonts,amsmath,bbm,bm,amssymb,mathtools}
\usepackage{xcolor}
\usepackage{ulem}
\usepackage{amsmath}
\usepackage[colorinlistoftodos]{todonotes}
\usepackage{caption} 
\usepackage[ruled,vlined]{algorithm2e}
\usepackage{xr}
\makeatletter
\newcommand*{\addFileDependency}[1]{
  \typeout{(#1)}
  \@addtofilelist{#1}
  \IfFileExists{#1}{}{\typeout{No file #1.}}
}
\makeatother

\newcommand*{\myexternaldocument}[1]{
    \externaldocument{#1}
    \addFileDependency{#1.tex}
    \addFileDependency{#1.aux}
}

\myexternaldocument{supp}

\def\Var{\text{Var}}

\def\log{\text{log}}

\def\E{\mathbb{E}}

\def\({\left(}
\def\){\right)}
\def\[{\left[}
\def\]{\right]}

\title{Efficient Penalized Generalized Linear Mixed Models for Variable Selection and Genetic Risk Prediction in High-Dimensional Data}

\author{JULIEN ST-PIERRE\thanks{To whom correspondence should be addressed.}\\[4pt]
\textit{Department of Epidemiology, Biostatistics and Occupational Health,} \\
\textit{McGill University, Montreal, Quebec, Canada}
\\[2pt]
{julien.st-pierre@mail.mcgill.ca}\\[4pt]
KARIM OUALKACHA\\[4pt]
\textit{Département de Mathématiques,} \\
\textit{Université du Québec à Montréal, Montreal, Quebec, Canada}\\[4pt]
SAHIR RAI BHATNAGAR\\[4pt]
\textit{Department of Epidemiology, Biostatistics and Occupational Health,} \\
\textit{McGill University, Montreal, Quebec, Canada}}

\begin{document}

\maketitle

\begin{abstract}
Sparse regularized regression methods are now widely used in genome-wide association studies (GWAS) to address the multiple testing burden that limits discovery of potentially important predictors. Linear mixed models (LMMs) have become an attractive alternative to principal components (PC) adjustment to account for population structure and relatedness in high-dimensional penalized models. However, their use in binary trait GWAS rely on the invalid assumption that the residual variance does not depend on the estimated regression coefficients. Moreover, LMMs use a single spectral decomposition of the covariance matrix of the responses, which is no longer possible in generalized linear mixed models (GLMMs). We introduce a new method called \texttt{pglmm}, a penalized GLMM that allows to simultaneously select genetic markers and estimate their effects, accounting for between-individual correlations and binary nature of the trait. We develop a computationally efficient algorithm based on PQL estimation that allows to scale regularized mixed models on high-dimensional binary trait GWAS ($\sim300,000$ SNPs). We show through simulations that penalized LMM and logistic regression with PC adjustment fail to correctly select important predictors and/or that prediction accuracy decreases for a binary response when the dimensionality of the relatedness matrix is high compared to \texttt{pglmm}. Further, we demonstrate through the analysis of two polygenic binary traits in the UK Biobank data that our method can achieve higher predictive performance, while also selecting fewer predictors than a sparse regularized logistic lasso with PC adjustment. Our method is available as a Julia package \texttt{PenalizedGLMM.jl}.
\end{abstract}

\section{Introduction} \label{sec:intro}

Genome-wide association studies (GWAS) have led to the identification of hundreds of common genetic variants, or single nucleotide polymorphisms (SNPs), associated with complex traits~\citep{Visscher2017} and are typically conducted by testing association on each SNP independently. However, these studies are plagued with the multiple testing burden that limits discovery of potentially important predictors. Moreover, GWAS have brought to light the problem of missing heritability, that is, identified variants only explain a low fraction of the total observed variability for traits under study~\citep{Manolio2009}. Multivariable regression methods, on the other hand, simultaneously fit many SNPs in a single model and are exempt from the multiple testing burden. In both simulations and analysis of high-dimensional data, sparse regularized logistic models have shown to achieve lower false-positive rates and higher precision than methods based on univariable GWAS summary statistics in case-control studies~\citep{Hoggart2008, Priv2019}. Contrary to univariable methods which implicitly assume that SNPs are independent, a regularized model makes use of the linkage disequilibrium (LD) structure between different loci, assigning weights to SNPs based on their relative importance after accounting for all other SNPs already in the model.

Confounding due to population structure or subject relatedness is another major issue in genetic association studies. Modern large scale cohorts will often include participants from different ethnic groups as well as admixed individuals, that is, subjects with individual-specific proportions of ancestries, or individuals with known or unknown familial relatedness, defined as cryptic relatedness~\citep{Sul2018}. Confounding comes from the fact that allele frequencies can differ greatly between individuals who do not share similar ancestry. When ignored, population structure and subject relatedness can decrease power and lead to spurious associations~\citep{Astle2009, Price2010}. Common practice is still to drop samples by applying filters for relatedness or genetic ancestry, which can result in decreasing the sample size by nearly 30\%~\citep{Loh2018} in the full UK Biobank data set~\citep{Bycroft2018}. 

Principal component analysis (PCA) can control for the confounding effect due to population structure by including the top eigenvectors of a genetic similarity matrix (GSM) as fixed effects in the regression model~\citep{Price2006}. 
With admixture and population structure being low dimensional fixed-effects processes, they can correctly be accounted for by using a relatively small number of PCs (e.g. 10)~\citep{Astle2009, Novembre2008}. However, using too few PCs can result in residual bias leading to false positives, while adding too many PCs as covariates can lead to a loss of efficiency~\citep{Zhao2018}. Alternatively, using mixed models (MMs), one can model population structure and/or closer relatedness by including a polygenic random effect with variance-covariance structure proportional to the GSM~\citep{Yu2005}. Indeed, kinship is a high-dimensional process, such that it cannot be fully captured by a few PCs~\citep{Hoffman2013}. Thus, it would require the inclusion of too many PCs as covariates, relative to the dimension of the sample size. Hence, while both PCA and MMs share the same underlying model, MMs are more robust in the sense that they do not require distinguishing between the different types of confounders~\citep{Price2010}. Moreover, MMs alleviate the need to evaluate the optimal number of PCs to retain in the model as fixed effects.

Several authors have proposed to combine penalized quasi-likelihood (PQL) estimation with sparsity inducing regularization to perform selection of fixed and/or random effects in generalized linear mixed model (GLMMs)~\citep{grollVariableSelectionGeneralized2014, huiJointSelectionMixed2017}. However, none of these methods are currently scalable for modern large-scale genome-wide data, nor can they directly incorporate relatedness structure through the use of a kinship matrix. Indeed, 
the computational efficiency of recent multivariable methods for high-dimensional MMs rely on performing a spectral decomposition of the covariance matrix to rotate the phenotype and design matrix such that the transformed data become uncorrelated~\citep{Bhatnagar2020, Rakitsch2012}. These methods are typically restricted to linear models since in GLMMs, it is no longer possible to perform a single spectral decomposition to rotate the phenotype and design matrix, as the covariance matrix depends on the sample weights which in turn depend on the estimated regression coefficients that are being iteratively updated. This limits the application of high-dimensional MMs to analysis of binary traits in genetic association studies.

In this paper, we introduce a new method called \texttt{pglmm} that allows to simultaneously select variables and estimate their effects, accounting for between-individual correlations and binary nature of the trait. We develop a scalable algorithm based on PQL estimation which makes it possible, for the first time, to fit penalized GLMMs on  high-dimensional GWAS data. To speedup computation, we estimate the variance components and dispersion parameter of the model under the null hypothesis of no genetic effect. Secondly, we use an upper-bound for the inverse variance-covariance matrix in order to perform a single spectral decomposition of the GSM and greatly reduce memory usage. Finally, we implement an efficient block coordinate descent algorithm in order to find the optimal estimates for the fixed and random effects parameters. Our method is implemented in an open source \texttt{Julia} programming language ~\citep{bezanson2017julia} package called \texttt{PenalizedGLMM.jl} and freely available at \url{https://github.com/julstpierre/PenalizedGLMM}.

The rest of this paper is structured as follows. In Section \ref{sec:methods} we present our model and describe the block coordinate gradient descent algorithm that is used to estimate the model parameters. We also discuss several approaches to select the optimal tuning parameter in regularized models, and we detail how predictions are obtained in GLMs with PC adjustment versus our proposed mixed model. In Section \ref{sec:results}, we show through simulations that both LMM and logistic model with PC adjustment fail to correctly select important predictors and estimate their effects when the dimensionality of the kinship matrix is high. Further, we demonstrate through the analysis of two polygenic binary traits in the UKBB data that our method achieves higher predictive performance, while also selecting consistently fewer predictors than a logistic lasso with PC adjustment. We finish with a discussion of our results, some limitations and future directions in Section \ref{sec:discussion}.

\section{Methods}\label{sec:methods}

\subsection{Model}\label{subsec:model}

We consider the following GLMM
\begin{align}\label{eq:model}
g(\mu_i) = \eta_i = \bm{X_}i\bm\alpha +\bm{G_}i\bm{\gamma} + b_i,
\end{align}
for $i=1,..,n$, where $\mu_i=\E(y_{i}=1 | \bm{X}_i, \bm{G}_i, b_i)$, $\bm{X}_i$ is a $1\times m$ row vector of covariates for subject $i$, $\bm\alpha$ is a $m \times 1$ column vector of fixed covariate effects including the intercept, $\bm{G}_i$ is a $1 \times p$ row vector of genotypes for subject $i$ taking values $\{0,1,2\}$ as the number of copies of the minor allele, and $\bm\gamma$ is a $p \times 1$ column vector of fixed additive genotype effects. We assume that $\bm{b}=(b_1,...,b_n)^\intercal \sim \mathcal{N}(0, \sum_{s=1}^S \tau_s \bm{V}_s)$ is an $n \times 1$ column vector of random effects, $\bm{\tau}=(\tau_1,...,\tau_S)^\intercal$ are variance component parameters, and $\bm{V}_s$ are known $n\times n$ relatedness matrices. The phenotypes $y_i$ are assumed to be conditionally independent and identically distributed given $(\bm{X}_i, \bm{G}_i, \bm{b})$ and follow any exponential family distribution with canonical link function $g(\cdot)$, mean $\E(y_i | \bm{b}) =\mu_i$ and variance $\Var(y_i| \bm{b}) = \phi a_i^{-1} \nu(\mu_i),$ where $\phi$ is a dispersion parameter, $a_i$ are known weights and $\nu(\cdot)$ is the variance function. In order to estimate the parameters of interest and perform variable selection, we need to use an approximation method to obtain a closed analytical form for the marginal likelihood of model \eqref{eq:model}. Following the derivation of~\citet{Chen2016}, we propose to fit  \eqref{eq:model} using a penalized quasi-likelihood (PQL) method, from where the log integrated quasi-likelihood function is equal to
\begin{align}\label{eq:A3}
ql(\bm{\alpha}, \bm{\gamma}, \phi, \bm{\tau}) = -\frac{1}{2}\text{log}\left|\sum_{s=1}^S\tau_s\bm{V}_s\bm{W} + \bm{I}_n\right| + \sum_{i=1}^n ql_i(\bm{\alpha}, \bm{\gamma}|\bm{\tilde{b}}) - \frac{1}{2}\bm{\tilde{b}}^\intercal \left(\sum_{s=1}^S \tau_s\bm{V}_s\right)^{-1}\bm{\tilde{b}},
\end{align}
where $\bm{W} = \textrm{diag}\left\{ \frac{a_i}{\phi\nu(\mu_i)[g'(\mu_i)^2]}\right\}$ is a diagonal matrix containing weights for each observation, $ql_i(\bm{\alpha,\gamma}|\bm{b}) = \int_{y_i}^{\mu_i}\frac{a_i(y_i-\mu)}{\phi\nu(\mu)} d\mu$ is the quasi-likelihood for the $ith$ individual given the random effects $\bm b$, and $\tilde{\bm{b}}$ is the solution which maximizes  \eqref{eq:A3}.

In typical genome-wide studies, the number of predictors is much greater than the number of observations ($p > n$), and the parameter vector $\bm{\gamma}$ becomes underdetermined when modelling all SNPs jointly. Thus, we propose to add a lasso regularization term~\citep{lasso} to the negative quasi-likelihood function in \eqref{eq:A3} to seek a sparse subset of $\bm{\gamma}$ that gives an adequate fit to the data. Because $ql(\bm{\alpha}, \bm{\gamma}, \phi, \bm{\tau})$ is a non-convex loss function, we propose a two-step estimation method to reduce the computational complexity. First, we obtain the variance component estimates $\hat{\phi}$ and $\bm{\hat{\tau}}$ under the null hypothesis of no genetic effect ($\bm{\gamma} = \bm{0}$) using the AI-REML algorithm~\citep{Gilmour1995, Chen2016} detailed in Appendix A of the supplementary material. Assuming that the weights in $\bm{W}$ vary slowly with the conditional mean, we drop the first term in \eqref{eq:A3}~\citep{breslowApproximateInferenceGeneralized1993} and define the following objective function which we seek to minimize with respect to $(\bm{\alpha}, \bm{\gamma}, \tilde{\bm{b}})$:
\begin{align}\label{eq:objfunc}
(\hat{\bm{\alpha}}, \hat{\bm{\gamma}}, \hat{\bm{b}}) &=  \underset{\bm{\alpha}, \bm{\gamma}, \tilde{\bm{b}}}{\text{argmin }} Q_{\lambda}(\bm{\alpha}, \bm{\gamma}, \tilde{\bm{b}}), \nonumber \\
Q_{\lambda}(\bm{\alpha}, \bm{\gamma}, \tilde{\bm{b}}) &= -\sum_{i=1}^n ql_i(\bm{\alpha}, \bm{\gamma}|\bm{\tilde{b}}) + \frac{1}{2}\bm{\tilde{b}}^\intercal\left(\sum_{s=1}^S \hat{\tau}_s\bm{V}_s\right)^{-1}\bm{\tilde{b}} + \lambda\sum_j v_j|\gamma_j| \nonumber \\
&:= -\ell_{PQL}(\bm{\alpha}, \bm{\gamma}, \hat\phi, \hat{\bm{\tau}}) +\lambda\sum_j v_j|\gamma_j|,
\end{align}
where $\lambda$ is a nonnegative regularization parameter, and $v_j$ is a penalty factor for the $j^{th}$ predictor.

In Appendix B, we detail our proposed general purpose block coordinate gradient descent algorithm (CGD) to solve \eqref{eq:objfunc} and obtain regularized PQL estimates for $\bm{\alpha},\bm{\gamma}$ and $\tilde{\bm{b}}$. Briefly, our algorithm is equivalent to iteratively solve the two penalized generalized least squares (GLS) $$\underset{\tilde{\bm{b}}}{\textrm{argmin}}\left(\tilde{\bm{Y}} - \tilde{\bm{X}}{\bm{\beta}} - \tilde{\bm{b}}\right)^\intercal\bm{W}^{-1}\left(\tilde{\bm{Y}} - \tilde{\bm{X}}{\bm{\beta}} - \tilde{\bm{b}}\right) + \tilde{\bm{b}}^\intercal\left(\sum_{s=1}^S \hat{\tau}_s\bm{V}_s\right)^{-1}\tilde{\bm{b}},$$ and $$\underset{\bm{\beta}}{\textrm{argmin}}\left(\tilde{\bm{Y}} - \tilde{\bm{X}}\bm{\beta}\right)^\intercal\bm{\Sigma}^{-1}\left(\tilde{\bm{Y}} - \tilde{\bm{X}}\bm{\beta}\right) + \lambda\sum_j v_j|\beta_j|,$$ where $\bm{\Sigma}=\bm{W}^{-1} + \sum_{s=1}^S \hat{\tau}_s\bm{V}_s$ is the covariance matrix of the working response vector $\tilde{\bm{Y}}$, $\tilde{\bm{X}}=\left[\bm{X};\ \bm{G}\right]$ and $\bm{\beta}=(\bm{\alpha}^\intercal, \bm{\gamma}^\intercal)^\intercal $. We use the spectral decomposition of $\bm{\Sigma}$ to rotate $\tilde{\bm{Y}}$, $\tilde{\bm{X}}$ and $\tilde{\bm{b}}$ such that the
transformed data become uncorrelated. For binary data, because the covariance matrix $\bm{\Sigma}$ depends on the sample weights $\bm{W}$, we use an upper-bound on $\bm{\Sigma}^{-1}$ to ensure a single spectral decomposition is performed~\citep{bohningMonotonicityQuadraticapproximationAlgorithms1988}. By cycling through the coordinates and minimizing the objective function with respect to one parameter at a time, $\tilde{\bm{b}}$ can be estimated by fitting a generalized ridge-like model with a diagonal penalty matrix equal to the inverse of the eigenvalues of $\sum_{s=1}^S \hat{\tau}_s\bm{V}_s$. Then, conditional on $\tilde{\bm{b}}$, ${\bm{\beta}}$ is estimated by solving a weighed least squares (WLS) with a lasso regularization term. All calculations and algorithmic steps are detailed in Appendix B.

\subsection{Model selection}\label{subsec:modelselection}

Approaches to selecting the optimal tuning parameter in regularized models are of primary interest since in real data analysis, the underlying true model is unknown. A popular strategy is to select the value of the tuning parameter that minimizes out-of-sample prediction error, e.g., cross-validation (CV), which is asymptotically equivalent to the Akaike information criterion (AIC) \citep{Akaike1998,Yang2005}. While being conceptually attractive, CV becomes computationally expensive for very high-dimensional data. Moreover, in studies where the proportion of related subjects is important, either by known or cryptic relatedness, the CV prediction error is no longer an unbiased estimator of the generalization error~\citep{Rabinowicz2020}. Through simulation studies and real data analysis, ~\citet{Wang2020} found that LD and minor allele frequencies (MAF) differences between ancestries could explain between 70 and 80\% of the loss of relative accuracy of European-based prediction models in African ancestry for traits like body mass index and type 2 diabetes. Thus, there is no clear approach to how multiple admixed and/or similar populations should be split when using CV to minimize out-of-sample prediction error.


Alternatively, we can use the generalized information criterion (GIC) to choose the optimal value of the tuning parameter $\lambda$, defined as 
\begin{align}\label{eq:GIC}
\textrm{GIC}_{\lambda} = -2 \ell_{PQL} + a_n \cdot \hat{df}_{\lambda},
\end{align}
where $\ell_{PQL}$ is defined in  \eqref{eq:objfunc}, and $\hat{df}_{\lambda}=|\{1 \le k \le p : \hat{\beta}_k \ne 0 \}| + \textrm{dim}(\hat{\bm{\tau}})$ is the number of nonzero fixed-effects coefficients~\citep{Zou2007} plus the number of variance components. Special cases of the GIC include AIC ($a_n=2$) and the Bayesian information criterion (BIC)~\citep{Schwarz1978} ($a_n=\log(n)$).

\subsection{Prediction}\label{subsec:prediction}
It is often of interest in genetic association studies to make predictions on a new set of individuals, e.g., the genetic risk of developing a disease for a binary response or the expected outcome in the case of a continuous response. In what follows, we compare how predictions are obtained in \texttt{pglmm} versus a GLM with PC adjustment. 

\subsubsection{\texttt{pglmm}}\label{subsubsec:mixed}

For the sake of comparison with the GLM with PC adjustment, we suppose a sampling design where a single variance component is needed such that $\bm{b}\sim\mathcal{N}(\bm{0}, \tau_1\bm{V_1})$ where $\bm{V_1}$ is the GSM between $n$ subjects that are used to fit the GLMM \eqref{eq:model}. We iteratively fit on a training set the working linear mixed model $$\tilde{\bm{Y}} = \tilde{\bm{X}}\bm{\beta} + \bm{\bm{b}} + \bm{\epsilon},$$ where $\bm{\epsilon}=g'(\bm{\mu})(\bm{y}-\bm{\mu}) \sim \mathcal{N}(0, \bm{W}^{-1})$. Let $\tilde{\bm{Y}}_s$ be the latent working vector in a set of individuals with predictor set $\tilde{\bm{X}}_s$ that were not used in the model training, $n_s$ denote the number of observations in the testing set and $n$ the number of observations in the training set. Similar to~\citep{Bhatnagar2020}, we assume that the marginal joint distribution of $\tilde{\bm{Y}}_s$ and $\tilde{\bm{Y}}$ is multivariate Normal :
\begin{align*}
\begin{bmatrix} \tilde{\bm{Y}}_s \\ \tilde{\bm{Y}}\end{bmatrix} \sim \mathcal{N}\left(\begin{bmatrix} \tilde{\bm{X}}_s\bm{\beta} \\ \tilde{\bm{X}}\bm{\beta}\end{bmatrix},\begin{bmatrix} \bm{\Sigma}_{11} & \bm{\Sigma}_{12} \\ \bm{\Sigma}_{21} & \bm{\Sigma}_{22} \end{bmatrix}\right),
\end{align*}
where $\bm{\Sigma}_{12}=\tau_1\bm{V}_{12}$ and $\bm{V}_{12}$ is the $n_s \times n$ GSM between the testing and training individuals. It follows from standard normal theory that
\begin{align*}
\tilde{\bm{Y}}_s|\tilde{\bm{Y}}, \phi,\tau_1, \bm{\beta}, \tilde{\bm{X}}, \tilde{\bm{X}}_s \sim \mathcal{N}\left(\tilde{\bm{X}}_s\bm{\beta} + \bm{\Sigma}_{12}\bm{\Sigma}_{22}^{-1}(\tilde{\bm{Y}}-\tilde{\bm{X}}\bm{\beta}), \bm{\Sigma}_{11}-\bm{\Sigma}_{12}\bm{\Sigma}_{22}^{-1}\bm{\Sigma}_{21}\right).
\end{align*}
The estimated mean response $\hat{{\bm{\mu}}}_s$ for the testing set is given by
\begin{align}\label{eq:predmix}
g^{-1}\left(\E[\tilde{\bm{Y}}_s|\tilde{\bm{Y}}, \hat\phi,\hat\tau_1, \hat{\bm{\beta}}, \tilde{\bm{X}}, \tilde{\bm{X}}_s]\right) &= g^{-1}\left(\tilde{\bm{X}}_s\hat{\bm{\beta}} + \bm{\Sigma}_{12}\bm{\Sigma}_{22}^{-1}(\tilde{\bm{Y}}-\tilde{\bm{X}}\hat{\bm{\beta}})\right) \nonumber \\
&= g^{-1}\left(\tilde{\bm{X}}_s\hat{\bm{\beta}} + \hat\tau_1\bm{V}_{12}\left(\bm{W}^{-1} + \hat\tau_1\bm{V}_1\right)^{-1}(\tilde{\bm{Y}}-\tilde{\bm{X}}\hat{\bm{\beta}})\right) \nonumber \\
&= g^{-1}\left(\tilde{\bm{X}}_s\hat{\bm{\beta}} + \bm{V}_{12}\bm{U}\left(\frac{1}{\hat\tau_1}\bm{D} + \tilde{\bm{U}}^\intercal\bm{W}\tilde{\bm{U}}\right)^{-1}\tilde{\bm{U}}^\intercal\bm{W}(\tilde{\bm{Y}}-\tilde{\bm{X}}\hat{\bm{\beta}})\right),
\end{align}
where $g(\cdot)$ is a link function and $\tilde{\bm{U}} = \bm{UD}$ is the $n \times n$ matrix of PCs obtained from the spectral decomposition of the GSM for training subjects. 

\subsubsection{GLM with PC adjustment}\label{subsubsec:glmPC}
Another approach to control for population structure and/or subjects relatedness is to use the first $r$ columns of $\tilde{\bm{U}}$ as unpenalized fixed effects covariates. This leads to the following GLM
\begin{align*}
g(\bm{\mu}) = \tilde{\bm{X}}\bm{\beta} + \tilde{\bm{U}}_r\bm{\delta},
\end{align*}
where $\tilde{\bm{X}}=[\bm{X}; \bm{G}]$ is the $n \times (m+p)$ design matrix for non-genetic and genetic predictors, $\bm{\beta} \in \mathbb{R}^p$ is the corresponding sparse vector of fixed effects, $\tilde{\bm{U}}_r$ is the $n \times r$ design matrix for the first $r$ PCs and $\delta \in \mathbb{R}^r$ is the corresponding vector of fixed effects. Letting  $\tilde{\bm{Y}} = \tilde{\bm{X}}\bm{\beta} + \tilde{\bm{U}}_r\bm{\delta} + g'(\bm{\mu})(\bm{y-\mu})$ be the working response vector, one can show that 
\begin{align}\label{eq:glspc}
\hat{\bm{\delta}} = \left(\tilde{\bm{U}}_r^\intercal\bm{W}\tilde{\bm{U}}_r\right)^{-1}\tilde{\bm{U}}_r^\intercal\bm{W}\left(\tilde{\bm{Y}} - \tilde{\bm{X}}\hat{\bm{\beta}}\right),
\end{align} 
where $\bm{W}$ is the diagonal matrix of GLM weights. Let $\bm{V}_{12}$ be the $n_s \times n$ GSM between a testing set of $n_s$ individuals and $n$ training individuals such that the projected PCs on the testing subjects are equal to $\bm{V}_{12}\bm{U}_{r}$. Then, the estimated mean response $\hat{{\bm{\mu}}}_s$ for the testing set is given by 
\begin{align}\label{eq:predpca}
\hat{\bm{\mu}}_s = g^{-1}\left(\tilde{\bm{X}_s}\hat{\bm{\beta}} + \bm{V}_{12}\bm{U}_{r}\hat{\bm{\delta}}\right) = g^{-1}\left(\tilde{\bm{X}_s}\hat{\bm{\beta}} + \bm{V}_{12}\bm{U}_{r}\left(\tilde{\bm{U}}_r^\intercal\bm{W}\tilde{\bm{U}}_r\right)^{-1}\tilde{\bm{U}}_r^\intercal\bm{W}\left(\tilde{\bm{Y}} - \tilde{\bm{X}}\hat{\bm{\beta}}\right)\right).
\end{align}

By comparing \eqref{eq:predmix} and \eqref{eq:predpca}, we see that both GLM with PC adjustment and \texttt{pglmm} use a projection of the training PCs on the testing set to predict new responses, but with different coefficients for the projected PCs. For the former, the estimated coefficients for the first $r$ projected PCs in \eqref{eq:glspc} are obtained by iteratively solving generalized least squares (GLS) on the partial working residuals $\tilde{\bm{Y}} - \tilde{\bm{X}}\hat{\bm{\beta}}$. For \texttt{pglmm}, the estimated coefficients for all projected PCs are also obtained by iteratively solving GLS on the partial working residuals $\tilde{\bm{Y}} - \tilde{\bm{X}}\hat{\bm{\beta}}$, with an extra ridge penalty for each coefficient that is equal to $\hat{\tau_1}^{-1}\Lambda_i$ with $\Lambda_i$ the $i^{th}$ eigenvalue of $\bm{V}$ that is associated with the $i^{th}$ PC.

From a Bayesian point of view, the fixed effect GLM assumes that each of the $r$ selected PCs have equal prior probability, while the remaining $n-r$ components are given a prior with unit mass at zero~\citep{Astle2009}. In contrast, the MM puts a Gaussian prior on regression coefficients with variances proportional to the corresponding eigenvalues. This implies that the PCs with the largest eigenvalues have a higher prior probability of explaining the phenotype. Hence, \texttt{pglmm} shrinks PCs coefficients in a smooth way, while the fixed effect GLM uses a tresholding approach; the first $r$ predictors with larger eigenvalues are kept intact, and the others are completely removed. This implies that the confounding effect from population structure and/or relatedness on the phenotype is fully captured by the first $r$ PCs. As we show in simulations results, departure from this assumption leads to less accurate coefficients estimation, lower prediction accuracy and higher variance in predictions.

\subsection{Simulation design}\label{subsec:simdesign}

We evaluated the performance of our proposed method against that of a lasso LMM, using the \texttt{R} package \texttt{ggmix}~\citep{ggmix}, and a logistic lasso, using the \texttt{Julia} package \texttt{GLMNet} which wraps the \texttt{Fortran} code from the original \texttt{R} package \texttt{glmnet}~\citep{glmnet}. We compared \texttt{glmnet} when we included or not the first 10 PCs in the model (\texttt{glmnetPC}). We performed a total of 50 replications for each simulation scenario, drawing anew genotypes and simulated traits. Values for all simulation parameters are presented in Table \ref{tab:1}.

\subsubsection{Simulated genotype from the admixture model}
In the first scenario, we studied the performance of all methods for different population structures by simulating random genotypes from the BN-PSD admixture model for 10 or 20 subpopulations with 1D geography or independent subpopulations using the \texttt{bnpsd} package in \texttt{R}~\citep{Ochoa2016_1, Ochoa2016_2}. Sample size was set to $n=2500$. We simulated $p$ candidate SNPs and randomly selected $p\times c$ to be causal. The kinship matrix $\bm{V}$ and PCs were calculated using a set of $50,000$ additional simulated SNPs. We simulated covariates for age and sex using Normal and Binomial distributions, respectively.

\subsubsection{Real genotypes from the UK Biobank data}
In the second scenario, we compared the performance of all methods when a high proportion of related individuals are present, using real genotype data from the UK Biobank. We retained a total of 6731 subjects of White British ancestry having estimated 1st, 2nd or 3rd degree relationships with at least one other individual. We sampled $p$ candidate SNPS among all chromosomes and randomly selected $p\times c$ to be causal. We used PCs as provided with the data set. These were computed using a set of unrelated samples and high quality markers pruned to minimise LD~\citep{Bycroft2018}. Then, all subjects were projected onto the principal components using the corresponding loadings. Since the markers that were used to compute the PCs were potentially sampled as candidate causal markers in our simulations, we included all candidate SNPs in the set of markers used for calculating the kinship matrix $\bm{V}$. We simulated age using a Normal distribution and used the sex covariate as provided with the data.

\subsubsection{Simulation model}
The number of candidate SNPs and fraction of causal SNPs were set to $p=5000$ and $c=0.01$ respectively. Let $S$ be the set of candidate causal SNPs, with $|S|=p\times c$, then the causal SNPs fixed effects $\beta_j$ were generated from a Gaussian distribution $\mathcal{N}(0,h^2_g\sigma^2/|S|)$, where $h^2_g$ is the fraction of variance on the logit scale that is due to total additive genetic fixed effects. That is, we assumed the candidate causal markers explained a fraction of the total polygenic heritability, and the rest was explained by a random polygenic effect $b\sim\mathcal{N}(0,h^2_b\sigma^2\bm{V})$. We simulated a SNR equal to 1 for the fixed genetic effects ($h^2_g=50\%$) under strong random polygenic effects ($h^2_b=40\%$). Finally, we simulated a binary phenotype using a logistic link function
\begin{align}\label{eq:simlogit1}
\text{logit}(\pi) = \text{logit}(\pi_0) -\text{log}(1.3)\times Sex+\text{log}(1.05)Age/10+ \sum_{j\in S}\beta_j\cdot \widetilde{G}_{j} + b,
\end{align}
where the parameter $\pi_0$ was chosen to specify the prevalence under the null, and $\widetilde{G}_{j}$ is the $j^{th}$ column of the standardized genotype matrix $\tilde{g}_{ij}=(g_{ij}-2p_i)/\sqrt{2p_i(1-p_i)}$ and $p_i$ is the MAF. 

\subsubsection{Metric of comparison}\label{subsubsec:metrics}
For each replication, subjects were partitioned into training and test sets using an 80/20 ratio, ensuring all related individuals were assigned into the same set in the second scenario. Variable selection and coefficient estimation were performed on training subjects for all methods. We compared each method at a fixed number of predictors, ranging from 5 to 50 which corresponds to the number of true causal SNPs. Comparisons were based on three criteria: the ability to retrieve the causal predictors, measured by the true positive rate ${\textrm{TPR} = |\{1\le k \le p : \hat{\beta}_k \ne 0 \cap {\beta}_k \ne 0\}|/|\{1\le k \le p : \hat{\beta}_k \ne 0\}|}$; the ability to accurately estimate coefficients, measured by the root mean squared error ${\textrm{RMSE} = \sqrt{\frac{1}{p}\sum_{k=1}^p (\hat{\beta}_k - \beta_k)^2}}$; and the ability to predict outcomes in the test sets, measured by the area under the roc curve (AUC).

In addition, we evaluated the performance of our proposed method when using either AIC, BIC or cross-validation as model selection criteria, rather than fixing the number of predictors in the model. For this, subjects from the real UKBB data were randomly split into training (40\%), validation (30\%) and test (30\%) sets, again ensuring all related individuals were assigned into the same set. For cross-validation, the full lasso solution path was fitted on the training set, and the regularization parameter was obtained on the model which maximized AUC on the validation set. We compared methods performance on the basis of TPR, AUC on the test sets and RMSE. Additionally, we compared each model selection approach on the total number of predictors selected and on the model precision, which is defined as the proportion of selected predictors that are true positives. 

\subsection{Real data application} \label{subsec:realdata}

We used the real UK Biobank data set presented in Section \ref{subsec:simdesign} to illustrate the potential advantages of \texttt{pglmm} over logistic lasso with PC adjustment for constructing a PRS on two highly heritable binary traits, asthma and high cholesterol, in a set of related individuals. Asthma is a common respiratory disease characterized by inflammation and partial obstruction of the bronchi in the lungs that results in difficulty breathing~\citep{Anderson2008}. High cholesterol can form plaques and fatty deposits on the walls of the arteries, and thus prevent the blood to circulate to the heart and brain. It is one of the main controllable risk factors for coronary artery disease, heart attack and stroke~\citep{kathiresan_polymorphisms_2008}.

After filtering for SNPs with missing rate greater than $0.01$, MAF above $0.05$ and a p-value for the Hardy–Weinberg exact test above $10^{-6}$, a total of 320K genotyped SNPs were remaining. To better understand the contribution of the PRS for predicting asthma and high cholesterol, we fitted for each trait a null model with only age, sex, genotyping array and the first 10 PCs as main effects. For our proposed \texttt{pglmm} method, we did not include any PC since kinship is accounted for by a random effect. Finally, we compared with a logistic lasso in which the top 10 PCs were included as unpenalized covariates in addition to age, sex and genotyping array. To find the optimal regularization parameter for both methods, we split the subjects in training (60\%), validation (20\%) and test (20\%) sets for a total of 40 times. For each replication, the full lasso solution path was fitted on the training set, and the regularization parameter was obtained on the model which maximized AUC on the validation set. We compared mean prediction accuracy on the test sets as well as the median number of predictors included in all models. Finally, we also compared our method's performance when the best model was chosen using BIC on the training fit. 

\section{Results}\label{sec:results}

\subsection{Simulation results from the admixture model} \label{subsec:simresultsadmix}

\begin{figure}[!h]
\caption{Mean of 50 TPRs for the simulated genotype data. $K$ represents the number of intermediate subpopulations in the 1d linear admixture data (left panel), and the number of independent subpopulations in the independent data (right panel). }\label{fig:tpr_sim}
\centering
\includegraphics[width=\textwidth]{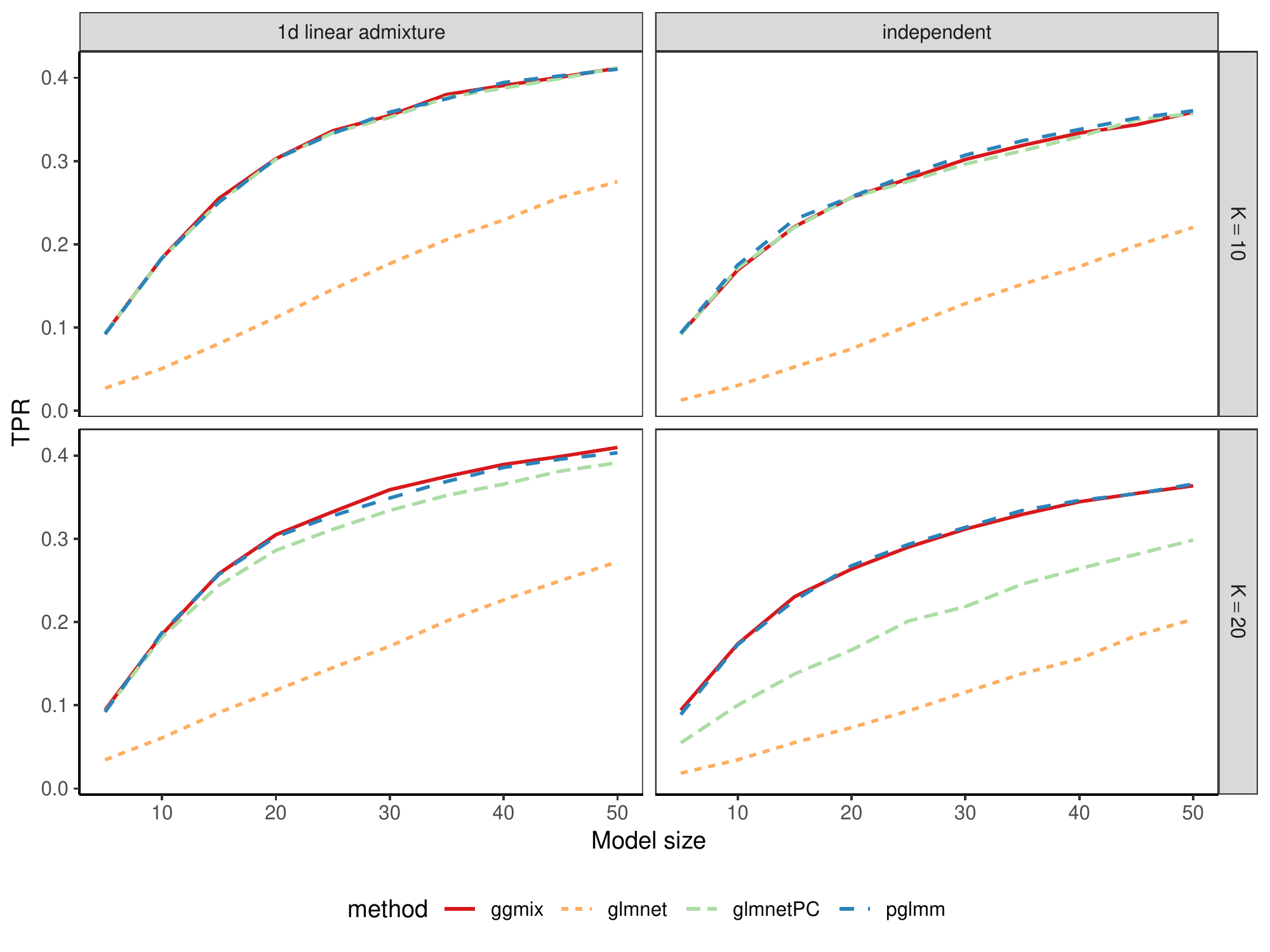}
\end{figure}

Results for selection of important predictors in the first simulation scenario, as measured by the mean TPR in 50 replications, are presented in Figure \ref{fig:tpr_sim}. For both 1D linear admixture and independent subpopulations, \texttt{glmnet} without PC adjustment failed to retrieve causal markers compared to all other methods. This is expected under population stratification; SNPs that differ in frequency between subpopulations are identified as important predictors because prevalence is not constant across each group. When the first 10 PCs were added as unpenalized covariates, \texttt{glmnetPC}'s ability to select causal predictors was lesser to that of \texttt{pglmm} and \texttt{ggmix} for the 20 independent subpopulations. In this case, the rank of the GSM used to infer the PCs is close to 20, and including only 10 PCs in the model does not correctly capture the confounding structure. Because there is less overlap between subpopulations in the admixture data compared to the independent populations~\citep{Reisetter2021}, a greater proportion of the simulated polygenic random effect is explained by the GSM and including only 10 PCs is enough to correct for confounding even when $K=20$ (bottom-left panel of Figure \ref{fig:tpr_sim}). On the other hand, including a random effect with variance-covariance structure proportional to the GSM correctly adjusts for population structure in all scenarios while alleviating the burden of choosing the right number of fixed predictors to include in the model. Even though \texttt{ggmix} assumes a standard LMM for the binary trait, it was able to identify causal markers at the same rate as \texttt{pglmm}. 

\begin{figure}[!h]
\caption{Mean of 50 RMSEs for the simulated genotype data. $K$ represents the number of intermediate subpopulations in the 1d linear admixture data (left panel), and the number of independent subpopulations in the independent data (right panel).}\label{fig:rmse_sim}
\centering
\includegraphics[width=\textwidth]{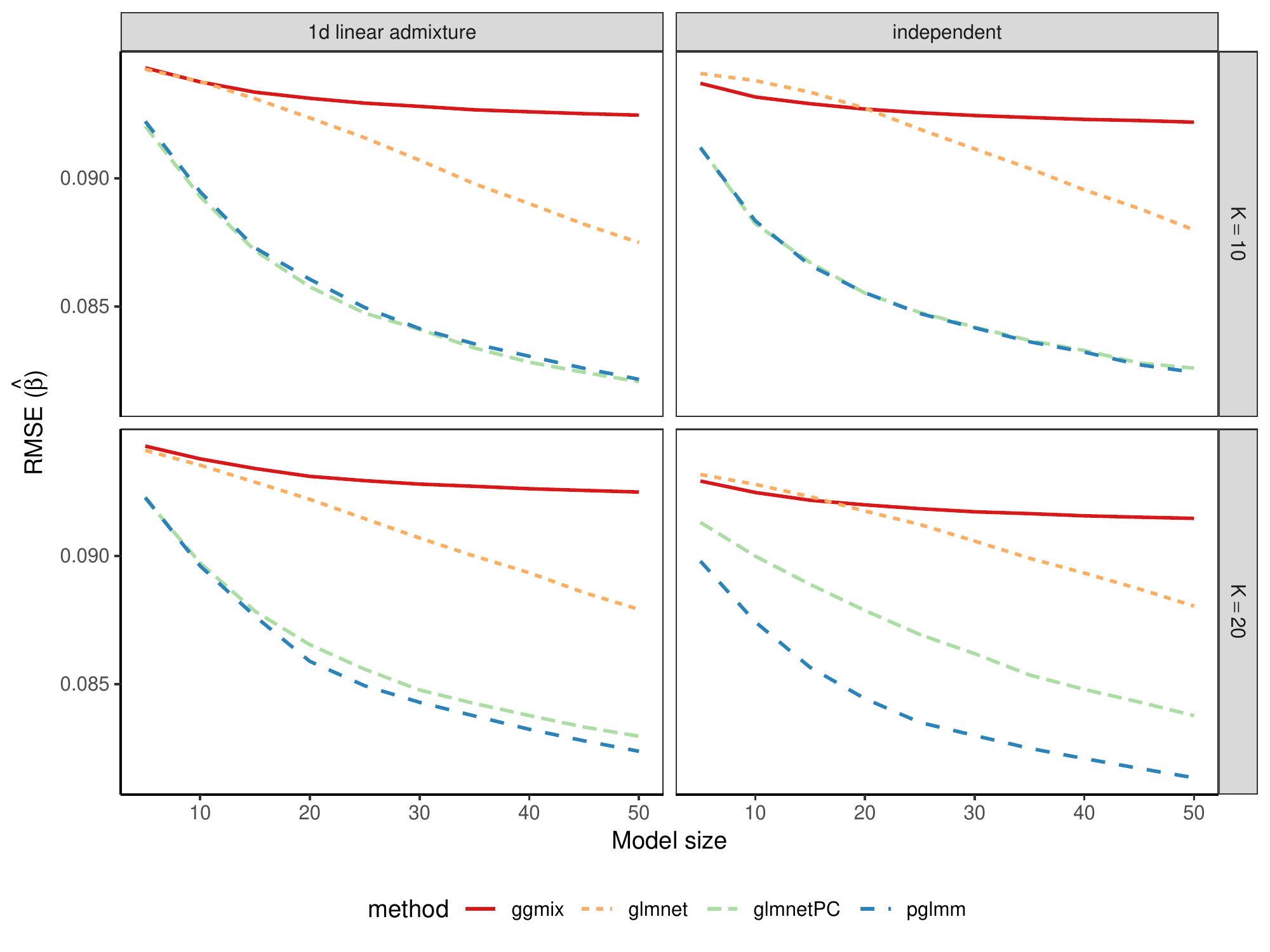}
\end{figure}

Results for estimation of SNP effects as measured by the mean RMSE in 50 replications are presented in Figure \ref{fig:rmse_sim}. Results are consistent with TPR results in that \texttt{glmnet} without PC adjustment performed poorly in all scenarios, while \texttt{pglmm} outperformed all other methods for the 20 independent subpopulations and performed comparably with \texttt{glmnetPC} for all other settings. As expected, \texttt{ggmix} had higher RMSE compared to \texttt{pglmm} and \texttt{glmnetPC}. Thus, even though \texttt{ggmix} was able to identify causal markers at the same rate as other methods that accounted for the binary nature of the response, resulting estimates for the SNP effects were not accurate. 

\begin{figure}[!h]
\caption{Mean of 50 AUCs in test sets of the simulated genotype data. $K$ represents the number of intermediate subpopulations in the 1d linear admixture data (left panel), and the number of independent subpopulations in the independent data (right panel). }\label{fig:auc_sim}
\centering
\includegraphics[width=\textwidth]{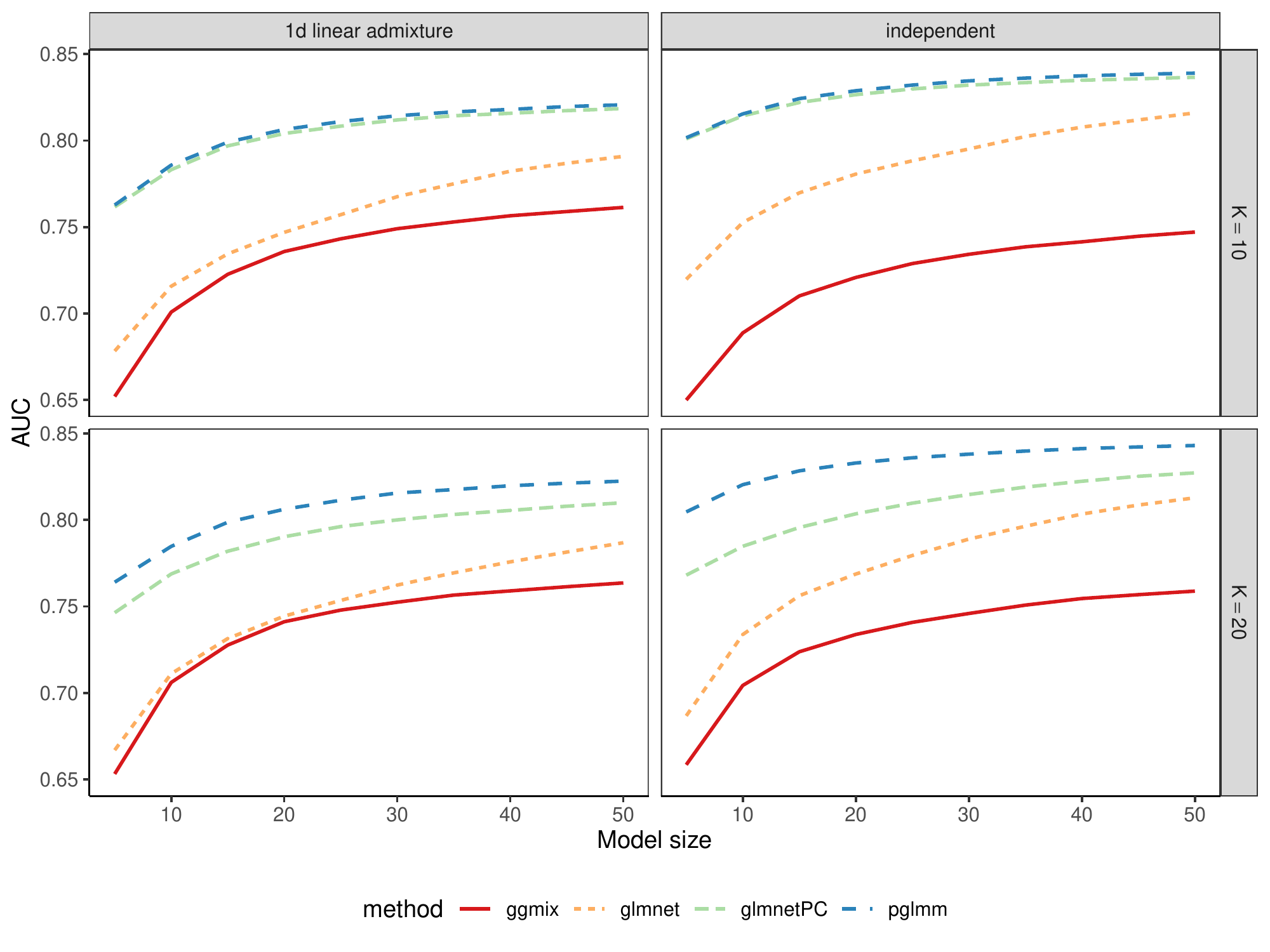}
\end{figure}

For both 1D linear admixture and independent subpopulations, \texttt{ggmix} and \texttt{glmnet} had poor predictive performance for $K=10$ and $K=20$, as reported in Figure \ref{fig:auc_sim}. Also, the predictive performance of \texttt{glmnetPC} was greatly reduced when $K=20$ for both admixture and independent populations. In the case of the admixture data, the RMSE for estimation of SNP effects was comparable for \texttt{glmnetPC} and \texttt{pglmm}. This means that the observed discrepancy in predictive accuracy is due to the difference in how each method handle the confounding effects. Using only 10 PCs as fixed effects when $K=20$ likely results in overfitted coefficients, which may potentially decrease prediction accuracy and increase variance of predictions in independent subjects. By using a ridge-like estimator for the random effects, $\texttt{pglmm}$ is less likely to overfit the confounding effects compared to \texttt{glmnetPC}. This is supported by the results of Table \ref{tab:3}, where the relative decrease in AUC standard deviation for the predictions obtained by $\texttt{pglmm}$ could be as high as $16\%$ for $K=20$ subpopulations. 

\subsection{Simulation results from real genotype data} \label{subsec:simresultsreal}

\begin{figure}[!h]
\caption{Mean of 50 AUCs, RMSEs and TPRs for the UK Biobank genotype data with related subjects.}\label{fig:ukbb}
\centering
\includegraphics[width=\textwidth]{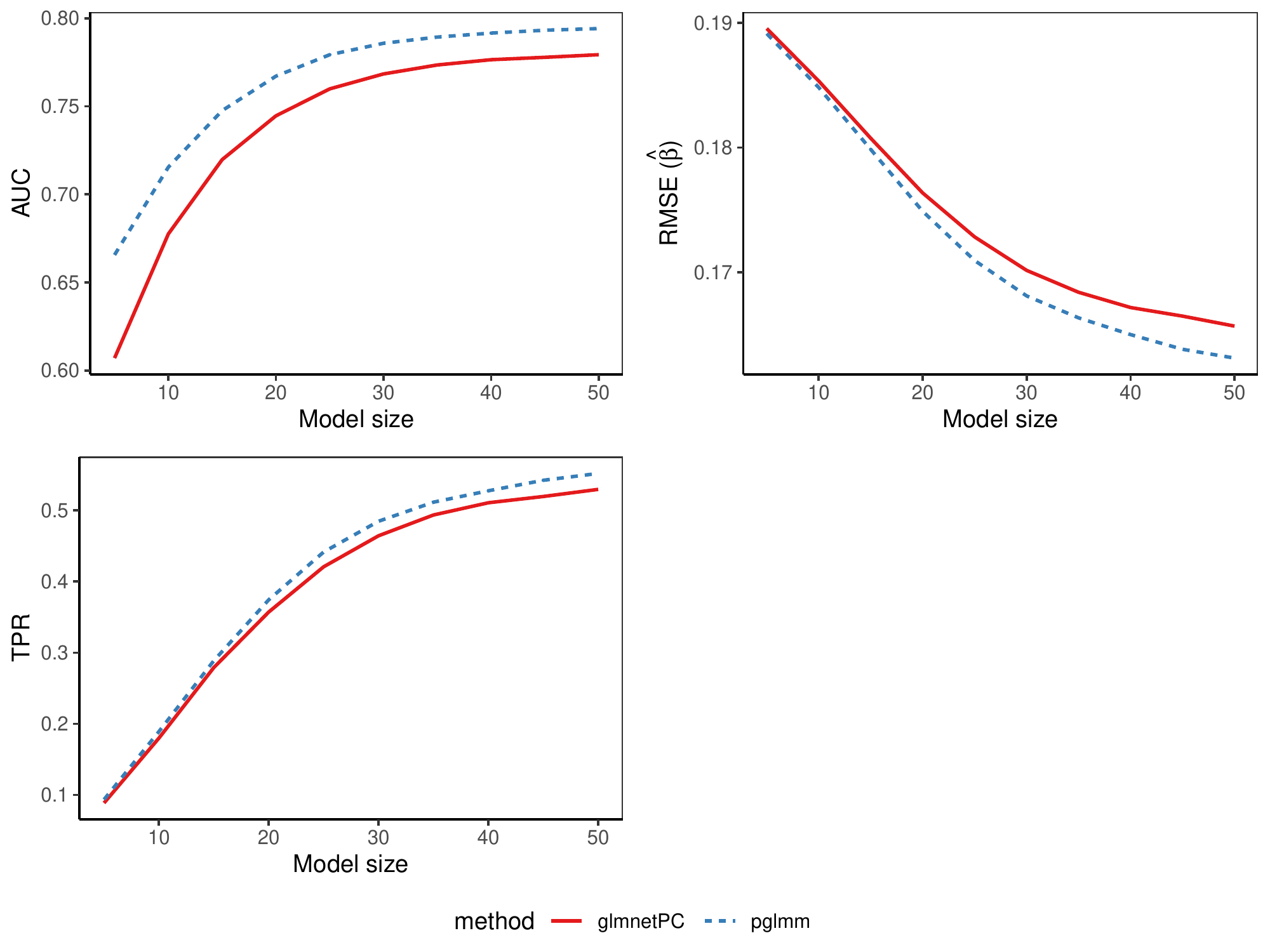}
\end{figure}

Results for selection of important predictors, estimation of SNPs effects and prediction accuracy in the second simulation scenario are presented in Figure \ref{fig:ukbb}. We compared the ability of \texttt{glmnetPC} and \texttt{pglmm} to adjust for potential confounding stemming from subjects relatedness. Both methods' ability to retrieve important predictors were comparable as measured by mean TPR, with \texttt{pglmm} having a slight advantage. In terms of predictor effect estimation, \texttt{pglmm} had lower reported mean RMSE. Furthermore, \texttt{pglmm} outperformed \texttt{glmnetPC} when making predictions in independent test sets. Once again, this is explained by the fact that \texttt{pglmm} uses a random effect parameterized by the $n-$dimensional kinship matrix, which we have shown in Section \ref{subsec:prediction} to be equivalent to include all PCs as predictors in the model and shrink their coefficients proportionally to their relative importance in a smooth way. On the other hand, for \texttt{glmnetPC}, only the first $10$ PCs with larger eigenvalues are kept intact, and the others are completely removed. As the confounding effect from relatedness on the phenotype can not be fully captured by using only the first $10$ PCs, prediction accuracy is greatly reduced.

\subsection{Model selection} \label{subsec:simresultsmodel}

\begin{figure}[!h]
\caption{Boxplots of the model selection simulations results for 50 replications of the UK Biobank genotype data with related subjects. For each replication, the best model for \texttt{pglmm} was chosen using either AIC, BIC or CV.}\label{fig:model_ukbb}
\centering
\includegraphics[width=\textwidth]{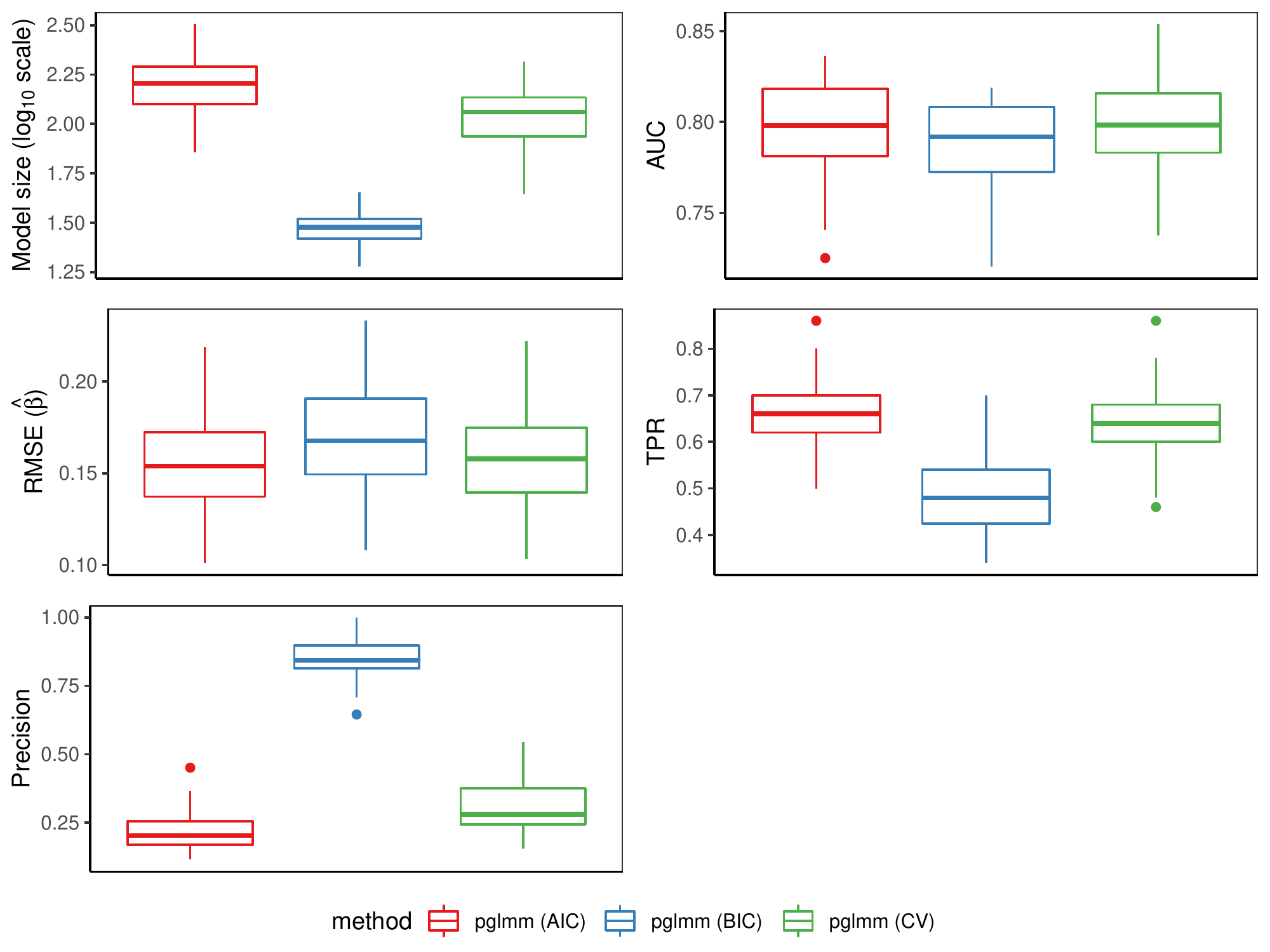}
\end{figure}

Boxplots of the model selection \sout{simulations} results are presented in Figure \ref{fig:model_ukbb}. As expected, BIC tended to choose sparser models with very high precision values, compared to AIC and CV which tended to select larger models with negligibly higher prediction performance. Thus, using BIC as a model selection criteria resulted in trading a bit of prediction accuracy for a large boost in the model precision. In many situations where it is of interest to identify a smaller set containing the most important predictors, BIC should be preferred over AIC and CV. Moreover, BIC alleviates the computational challenge of performing out-of-sample predictions, which includes identifying pedigrees to ensure independence between training, validation and testing sets.

\subsection{PRS for the UK Biobank} \label{subsec:realdataresults}

Results for asthma and high cholesterol PRSs are summarized in Table \ref{tab:2}. For asthma, \texttt{pglmm} with either BIC or CV as model selection criteria performed better than \texttt{glmnetPC} and the null model with covariates only when comparing mean AUC on the test sets. The median number of predictors selected by \texttt{glmnetPC} was four times higher than for \texttt{glmnet} when using CV for both methods. Moreover, the variability in predictors selected was more important for \texttt{glmnetPC}, as reported by an IQR value equal to 486, compared to 145 for our method. \texttt{pglmm} with BIC selected 1 predictor (IQR: 1) compared to 16 (IQR: 145) for \texttt{pglmm} with CV. This is consistent with our simulation results showing that BIC results in sparser models with comparable predictive power. For high cholesterol, very few genetic predictors were selected by all models, which suggests that it may not be a highly polygenic trait. In fact, using only the non-genetic covariates and first 10 PCs resulted in the best model \sout{for high cholesterol} based on mean test sets AUC.  

\section{Discussion}\label{sec:discussion}

We have introduced a new method called \texttt{pglmm} based on regularized PQL estimation, for selecting important predictors and estimating their effects in high-dimensional GWAS data, accounting for population structure, close relatedness and binary nature of the trait. Through a variety of simulations, using both simulated and real genotype data, we showed that \texttt{pglmm} was markedly better than a logistic lasso with PC adjustment when the number of subpopulations was greater than the number of PCs included, or when a high proportion of related subjects were present. We also showed that a lasso LMM was unable to estimate predictor effects with accuracy for binary responses, which greatly decreased its predictive performance. Performance assessment was based on TPR of selected predictors, RMSE of estimated effects and AUC of predictions. These results strongly advocate for using methods that explicitly account for the binary nature of the trait while effectively controlling for population structure and relatedness in genetic studies. 

When the dimensionality of the confounding structure was low, we showed that \texttt{pglmm} was equivalent to logistic lasso with PC adjustement. Hence, adjusting a GLM with PCA is at best equivalent to \texttt{pglmm}, but with the additional burden of selecting an appropriate number of PCs to retain in the model. Estimating the dimensionality of real datasets, and thus the number of PCs to include as fixed effects in a regression model can reveal to be challenging because estimated eigenvalues have biased distributions~\citep{Yao2022}. Another strategy involves selecting the appropriate number of PCs based on the Tracy-Widom test~\citep{Tracy1994}. However, it is known that this test tends to select a very large number of PCs~\citep{Lin2011}, causing convergence problems when fitting too many predictors. On the other hand, modeling the population structure by using a random polygenic effect correctly accounts for low and high-dimensional confounding structures, while only fitting one extra variance component parameter. 

We used real genotype data from the UK Biobank to simulate binary responses and showed that BIC effectively selected sparser models with very high precision and prediction accuracy, compared to AIC and CV. Using the same data set, we illustrated the potential advantages of \texttt{pglmm} over a logistic lasso with PC adjustment for constructing a PRS on two highly heritable binary traits in a set of related individuals. Results showed that \texttt{pglmm} had higher predictive performance for asthma, while also selecting consistently fewer predictors as reported by median and IQR values.

In this study, we focused solely on the lasso as a regularization penalty for the genetic markers effects. However, it is known that estimated effects by lasso will have large biases because the resulting shrinkage is constant irrespective of the magnitude of the effects. Alternative regularization like the Smoothly Clipped Absolute Deviation (SCAD)~\citep{Fan2001} penalty function should be explored. Although, we note that the number of nonzero coefficients in the SCAD estimates is no longer an unbiased estimate of its degrees of freedom. Other alternatives include implementation of the relaxed lasso, which has shown to produce sparser models with equal or lower prediction loss than the regular lasso estimator for high-dimensional data~\citep{Meinshausen2007}. It would also be of interest to explore if tuning of the generalized ridge penalty term on the random effects that arises from the PQL loss could result in better predictive performance. 

A limitation of \texttt{pglmm} compared to a logistic lasso with PC adjustment is the computational cost of performing multiple matrix calculations that comes from the estimation of variance components under the null. Indeed, at each iteration, we perform a matrix inversion based on Cholesky decomposition with complexity of $O(n^3)$ and matrix multiplications with complexity of $O(mn^2 + S^2 n^2 + p^2n)$, where $n$ is the sample size, $m$ is the number of non-genetic covariates, and $S$ is the number of variance components. Then, we need to perform a spectral decomposition of the covariance matrix with a computation time $O(n^3)$. These computations become prohibitive for large cohorts such as the full UK Biobank with a total of 500$K$ samples. A solution to explore to increase computation speed and decrease memory usage would be the use of conjugate gradient methods with a diagonal preconditioner matrix, as proposed by~\citet{zhouEfficientlyControllingCasecontrol2018}.

Finally, we can take advantage of the fact that it is possible to allow for multiple random effects to account for complex sampling designs and extend \texttt{pglmm} to a wide variety of models. For example, building a PRS for a bivariate binary trait, explicitly accounting for the shared causal pathways of many diseases or complex traits. Moreover, \texttt{pglmm} could be used in models where there is interest in selecting over fixed genetic and gene-environment interaction (GEI) effects. Due to the hierarchical structure between the main genetic and GEI effects, we will have to consider using a lasso for hierarchical structures~\citep{Zemlianskaia2022}.

\section*{Software}\label{sec:software}
Our Julia package called \texttt{PenalizedGLMM} is available on Github \url{https://github.com/julstpierre/PenalizedGLMM}.

\section*{Funding}
This work was supported by the Fonds de recherche Québec-Santé [267074 to K.O.]; and the Natural Sciences and Engineering Research Council of Canada [RGPIN-2019-06727 to K.O., RGPIN-2020-05133 to S.B.].

\section*{Acknowledgments}
This research has been conducted using the UK Biobank Resource under Application Number 20802.  This study was enabled in part by support provided by Calcul Québec (\url{https://www.calculquebec.ca/}) and Compute Canada (\url{https://www.computecanada. ca/}). We thank the UK Biobank and all participants for providing information.

\section*{Data availability statement}\label{sec:dataavail}

Simulated data are available on Github \url{https://github.com/julstpierre/PenalizedGLMM/data}. UK Biobank data are available via application directly to UK Biobank (\url{https://www.ukbiobank.ac.uk/enable-your-research}). The current study was conducted under UK Biobank application number 20802. 

\noindent{\it Conflict of Interest}: None declared.

\bibliographystyle{apalike}
\bibliography{biblio}

\begin{thebibliography}{}

\bibitem[Akaike, 1998]{Akaike1998}
Akaike, H. (1998).
\newblock {\em Information Theory and an Extension of the Maximum Likelihood
  Principle}, pages 199--213.
\newblock Springer New York, New York, NY.

\bibitem[Anderson, 2008]{Anderson2008}
Anderson, G.~P. (2008).
\newblock Endotyping asthma: new insights into key pathogenic mechanisms in a
  complex, heterogeneous disease.
\newblock {\em The Lancet}, 372(9643):1107--1119.

\bibitem[Astle and Balding, 2009]{Astle2009}
Astle, W. and Balding, D.~J. (2009).
\newblock Population structure and cryptic relatedness in genetic association
  studies.
\newblock {\em Statistical Science}, 24(4):451--471.

\bibitem[Bezanson et~al., 2017]{bezanson2017julia}
Bezanson, J., Edelman, A., Karpinski, S., and Shah, V.~B. (2017).
\newblock Julia: A fresh approach to numerical computing.
\newblock {\em SIAM review}, 59(1):65--98.

\bibitem[Bhatnagar et~al., 2020a]{ggmix}
Bhatnagar, S.~R., Yang, Y., and Greenwood, C. M.~T. (2020a).
\newblock ggmix: Variable selection in linear mixed models for snp data. {R}
  package version 0.0.1.

\bibitem[Bhatnagar et~al., 2020b]{Bhatnagar2020}
Bhatnagar, S.~R., Yang, Y., Lu, T., Schurr, E., Loredo-Osti, J., Forest, M.,
  Oualkacha, K., and Greenwood, C. M.~T. (2020b).
\newblock Simultaneous {SNP} selection and adjustment for population structure
  in high dimensional prediction models.
\newblock {\em {PLOS} Genetics}, 16(5):e1008766.

\bibitem[B{\"o}hning and Lindsay,
  1988]{bohningMonotonicityQuadraticapproximationAlgorithms1988}
B{\"o}hning, D. and Lindsay, B.~G. (1988).
\newblock Monotonicity of quadratic-approximation algorithms.
\newblock {\em Annals of the Institute of Statistical Mathematics},
  40(4):641--663.

\bibitem[Breslow and Clayton, 1993]{breslowApproximateInferenceGeneralized1993}
Breslow, N.~E. and Clayton, D.~G. (1993).
\newblock Approximate {{Inference}} in {{Generalized Linear Mixed Models}}.
\newblock {\em Journal of the American Statistical Association}, 88(421):9--25.

\bibitem[Bycroft et~al., 2018]{Bycroft2018}
Bycroft, C., Freeman, C., Petkova, D., Band, G., Elliott, L.~T., Sharp, K.,
  Motyer, A., Vukcevic, D., Delaneau, O., O'Connell, J., Cortes, A., Welsh, S.,
  Young, A., Effingham, M., McVean, G., Leslie, S., Allen, N., Donnelly, P.,
  and Marchini, J. (2018).
\newblock The {UK} biobank resource with deep phenotyping and genomic data.
\newblock {\em Nature}, 562(7726):203--209.

\bibitem[Chen et~al., 2016]{Chen2016}
Chen, H., Wang, C., Conomos, M.~P., Stilp, A.~M., Li, Z., Sofer, T., Szpiro,
  A.~A., Chen, W., Brehm, J.~M., Celed{\'{o}}n, J.~C., Redline, S.,
  Papanicolaou, G.~J., Thornton, T.~A., Laurie, C.~C., Rice, K., and Lin, X.
  (2016).
\newblock Control for population structure and relatedness for binary traits in
  genetic association studies via logistic mixed models.
\newblock {\em The American Journal of Human Genetics}, 98(4):653--666.

\bibitem[Fan and Li, 2001]{Fan2001}
Fan, J. and Li, R. (2001).
\newblock Variable selection via nonconcave penalized likelihood and its oracle
  properties.
\newblock {\em Journal of the American Statistical Association},
  96(456):1348--1360.

\bibitem[Friedman et~al., 2010]{glmnet}
Friedman, J., Hastie, T., and Tibshirani, R. (2010).
\newblock Regularization paths for generalized linear models via coordinate
  descent.
\newblock {\em Journal of Statistical Software}, 33(1):1--22.

\bibitem[Gilmour et~al., 1995]{Gilmour1995}
Gilmour, A.~R., Thompson, R., and Cullis, B.~R. (1995).
\newblock Average information {REML}: An efficient algorithm for variance
  parameter estimation in linear mixed models.
\newblock {\em Biometrics}, 51(4):1440.

\bibitem[Groll and Tutz, 2014]{grollVariableSelectionGeneralized2014}
Groll, A. and Tutz, G. (2014).
\newblock Variable selection for generalized linear mixed models by {{L}}
  1-penalized estimation.
\newblock {\em Statistics and Computing}, 24(2):137--154.

\bibitem[Hoffman, 2013]{Hoffman2013}
Hoffman, G.~E. (2013).
\newblock Correcting for population structure and kinship using the linear
  mixed model: Theory and extensions.
\newblock {\em {PLoS} {ONE}}, 8(10):e75707.

\bibitem[Hoggart et~al., 2008]{Hoggart2008}
Hoggart, C.~J., Whittaker, J.~C., Iorio, M.~D., and Balding, D.~J. (2008).
\newblock Simultaneous analysis of all {SNPs} in genome-wide and re-sequencing
  association studies.
\newblock {\em {PLoS} Genetics}, 4(7):e1000130.

\bibitem[Hui et~al., 2017]{huiJointSelectionMixed2017}
Hui, F. K.~C., M{\"u}ller, S., and Welsh, A.~H. (2017).
\newblock Joint {{Selection}} in {{Mixed Models}} using {{Regularized PQL}}.
\newblock {\em Journal of the American Statistical Association},
  112(519):1323--1333.

\bibitem[Kathiresan et~al., 2008]{kathiresan_polymorphisms_2008}
Kathiresan, S., Melander, O., Anevski, D., Guiducci, C., Burtt, N.~P., Roos,
  C., Hirschhorn, J.~N., Berglund, G., Hedblad, B., Groop, L., Altshuler,
  D.~M., Newton-Cheh, C., and Orho-Melander, M. (2008).
\newblock Polymorphisms {Associated} with {Cholesterol} and {Risk} of
  {Cardiovascular} {Events}.
\newblock {\em New England Journal of Medicine}, 358(12):1240--1249.

\bibitem[Lin and Zeng, 2011]{Lin2011}
Lin, D.~Y. and Zeng, D. (2011).
\newblock Correcting for population stratification in genomewide association
  studies.
\newblock {\em Journal of the American Statistical Association},
  106(495):997--1008.

\bibitem[Loh et~al., 2018]{Loh2018}
Loh, P.-R., Kichaev, G., Gazal, S., Schoech, A.~P., and Price, A.~L. (2018).
\newblock Mixed-model association for biobank-scale datasets.
\newblock {\em Nature Genetics}, 50(7):906--908.

\bibitem[Manolio et~al., 2009]{Manolio2009}
Manolio, T.~A., Collins, F.~S., Cox, N.~J., Goldstein, D.~B., Hindorff, L.~A.,
  Hunter, D.~J., McCarthy, M.~I., Ramos, E.~M., Cardon, L.~R., Chakravarti, A.,
  Cho, J.~H., Guttmacher, A.~E., Kong, A., Kruglyak, L., Mardis, E., Rotimi,
  C.~N., Slatkin, M., Valle, D., Whittemore, A.~S., Boehnke, M., Clark, A.~G.,
  Eichler, E.~E., Gibson, G., Haines, J.~L., Mackay, T. F.~C., McCarroll,
  S.~A., and Visscher, P.~M. (2009).
\newblock Finding the missing heritability of complex diseases.
\newblock {\em Nature}, 461(7265):747--753.

\bibitem[Meinshausen, 2007]{Meinshausen2007}
Meinshausen, N. (2007).
\newblock Relaxed lasso.
\newblock {\em Computational Statistics \& Data Analysis}, 52(1):374--393.

\bibitem[Novembre and Stephens, 2008]{Novembre2008}
Novembre, J. and Stephens, M. (2008).
\newblock Interpreting principal component analyses of spatial population
  genetic variation.
\newblock {\em Nature Genetics}, 40(5):646--649.

\bibitem[Ochoa and Storey, 2016a]{Ochoa2016_1}
Ochoa, A. and Storey, J.~D. (2016a).
\newblock {FST} and kinship for arbitrary population structures i: Generalized
  definitions.

\bibitem[Ochoa and Storey, 2016b]{Ochoa2016_2}
Ochoa, A. and Storey, J.~D. (2016b).
\newblock {FST} and kinship for arbitrary population structures {II}:
  Method-of-moments estimators.

\bibitem[Price et~al., 2006]{Price2006}
Price, A.~L., Patterson, N.~J., Plenge, R.~M., Weinblatt, M.~E., Shadick,
  N.~A., and Reich, D. (2006).
\newblock Principal components analysis corrects for stratification in
  genome-wide association studies.
\newblock {\em Nature Genetics}, 38(8):904--909.

\bibitem[Price et~al., 2010]{Price2010}
Price, A.~L., Zaitlen, N.~A., Reich, D., and Patterson, N. (2010).
\newblock New approaches to population stratification in genome-wide
  association studies.
\newblock {\em Nature Reviews Genetics}, 11(7):459--463.

\bibitem[Priv{\'{e}} et~al., 2019]{Priv2019}
Priv{\'{e}}, F., Aschard, H., and Blum, M. G.~B. (2019).
\newblock Efficient implementation of penalized regression for genetic risk
  prediction.
\newblock {\em Genetics}, 212(1):65--74.

\bibitem[Rabinowicz and Rosset, 2020]{Rabinowicz2020}
Rabinowicz, A. and Rosset, S. (2020).
\newblock Cross-validation for correlated data.
\newblock {\em Journal of the American Statistical Association}, pages 1--14.

\bibitem[Rakitsch et~al., 2012]{Rakitsch2012}
Rakitsch, B., Lippert, C., Stegle, O., and Borgwardt, K. (2012).
\newblock A lasso multi-marker mixed model for association mapping with
  population structure correction.
\newblock {\em Bioinformatics}, 29(2):206--214.

\bibitem[Reisetter and Breheny, 2021]{Reisetter2021}
Reisetter, A.~C. and Breheny, P. (2021).
\newblock Penalized linear mixed models for structured genetic data.
\newblock {\em Genetic Epidemiology}.

\bibitem[Schwarz, 1978]{Schwarz1978}
Schwarz, G. (1978).
\newblock Estimating the dimension of a model.
\newblock {\em The Annals of Statistics}, 6(2).

\bibitem[Sul et~al., 2018]{Sul2018}
Sul, J.~H., Martin, L.~S., and Eskin, E. (2018).
\newblock Population structure in genetic studies: Confounding factors and
  mixed models.
\newblock {\em {PLOS} Genetics}, 14(12):e1007309.

\bibitem[Tibshirani, 1996]{lasso}
Tibshirani, R. (1996).
\newblock Regression shrinkage and selection via the lasso.
\newblock {\em Journal of the Royal Statistical Society. Series B
  (Methodological)}, 58(1):267--288.

\bibitem[Tracy and Widom, 1994]{Tracy1994}
Tracy, C.~A. and Widom, H. (1994).
\newblock Level-spacing distributions and the airy kernel.
\newblock {\em Communications in Mathematical Physics}, 159(1):151--174.

\bibitem[Visscher et~al., 2017]{Visscher2017}
Visscher, P.~M., Wray, N.~R., Zhang, Q., Sklar, P., McCarthy, M.~I., Brown,
  M.~A., and Yang, J. (2017).
\newblock 10 years of {GWAS} discovery: Biology, function, and translation.
\newblock {\em The American Journal of Human Genetics}, 101(1):5--22.

\bibitem[Wang et~al., 2020]{Wang2020}
Wang, Y., Guo, J., Ni, G., Yang, J., Visscher, P.~M., and Yengo, L. (2020).
\newblock Theoretical and empirical quantification of the accuracy of polygenic
  scores in ancestry divergent populations.
\newblock {\em Nature Communications}, 11(1).

\bibitem[Yang, 2005]{Yang2005}
Yang, Y. (2005).
\newblock Can the strengths of aic and bic be shared? a conflict between model
  indentification and regression estimation.
\newblock {\em Biometrika}, 92(4):937--950.

\bibitem[Yao and Ochoa, 2022]{Yao2022}
Yao, Y. and Ochoa, A. (2022).
\newblock Limitations of principal components in quantitative genetic
  association models for human studies.

\bibitem[Yu et~al., 2005]{Yu2005}
Yu, J., Pressoir, G., Briggs, W.~H., Bi, I.~V., Yamasaki, M., Doebley, J.~F.,
  McMullen, M.~D., Gaut, B.~S., Nielsen, D.~M., Holland, J.~B., Kresovich, S.,
  and Buckler, E.~S. (2005).
\newblock A unified mixed-model method for association mapping that accounts
  for multiple levels of relatedness.
\newblock {\em Nature Genetics}, 38(2):203--208.

\bibitem[Zemlianskaia et~al., 2022]{Zemlianskaia2022}
Zemlianskaia, N., Gauderman, W.~J., and Lewinger, J.~P. (2022).
\newblock A scalable hierarchical lasso for gene-environment interactions.
\newblock {\em Journal of Computational and Graphical Statistics}, pages 1--36.

\bibitem[Zhao et~al., 2018]{Zhao2018}
Zhao, H., Mitra, N., Kanetsky, P.~A., Nathanson, K.~L., and Rebbeck, T.~R.
  (2018).
\newblock A practical approach to adjusting for population stratification in
  genome-wide association studies: principal components and propensity scores
  ({PCAPS}).
\newblock {\em Statistical Applications in Genetics and Molecular Biology},
  17(6).

\bibitem[Zhou et~al., 2018]{zhouEfficientlyControllingCasecontrol2018}
Zhou, W., Nielsen, J.~B., Fritsche, L.~G., Dey, R., Gabrielsen, M.~E., Wolford,
  B.~N., LeFaive, J., VandeHaar, P., Gagliano, S.~A., Gifford, A., Bastarache,
  L.~A., Wei, W.-Q., Denny, J.~C., Lin, M., Hveem, K., Kang, H.~M., Abecasis,
  G.~R., Willer, C.~J., and Lee, S. (2018).
\newblock Efficiently controlling for case-control imbalance and sample
  relatedness in large-scale genetic association studies.
\newblock {\em Nature Genetics}, 50(9):1335--1341.

\bibitem[Zou et~al., 2007]{Zou2007}
Zou, H., Hastie, T., and Tibshirani, R. (2007).
\newblock On the {\textquotedblleft}degrees of freedom{\textquotedblright} of
  the lasso.
\newblock {\em The Annals of Statistics}, 35(5).

\end{thebibliography}


\begin{thebibliography}{}

\bibitem[B{\"o}hning and Lindsay,
  1988]{bohningMonotonicityQuadraticapproximationAlgorithms1988}
B{\"o}hning, D. and Lindsay, B.~G. (1988).
\newblock Monotonicity of quadratic-approximation algorithms.
\newblock {\em Annals of the Institute of Statistical Mathematics},
  40(4):641--663.

\bibitem[Chen et~al., 2016]{Chen2016}
Chen, H., Wang, C., Conomos, M.~P., Stilp, A.~M., Li, Z., Sofer, T., Szpiro,
  A.~A., Chen, W., Brehm, J.~M., Celed{\'{o}}n, J.~C., Redline, S.,
  Papanicolaou, G.~J., Thornton, T.~A., Laurie, C.~C., Rice, K., and Lin, X.
  (2016).
\newblock Control for population structure and relatedness for binary traits in
  genetic association studies via logistic mixed models.
\newblock {\em The American Journal of Human Genetics}, 98(4):653--666.

\bibitem[Friedman et~al., 2007]{friedmanPathwiseCoordinateOptimization2007}
Friedman, J., Hastie, T., H{\"o}fling, H., and Tibshirani, R. (2007).
\newblock Pathwise coordinate optimization.
\newblock {\em The Annals of Applied Statistics}, 1(2).

\end{thebibliography}

\clearpage
\section*{Tables}\label{sec:tables}

\begin{table}[h]
\centering
\caption{Simulation parameters}\label{tab:1}
\begin{tabular}{l||l|c}
\hline
Parameter & Definition  & Value                                                        \\ \hline
$M$       & Number of replications & 50 \\
$h^2_g$ & Fraction of variance due to fixed & 0.5 \\ & genetic effects (logit scale) & \\
$h^2_b$ & Fraction of variance due to random & 0.4 \\ & genetic effects (logit scale) &  \\
$\pi_0$   & Prevalence under the null &  0.1 \\
$p$ & Number of snps & 5,000 \\
$c$       & Fraction of causal SNPs &   0.01 \\
\hline
\end{tabular}
\end{table}

\begin{table}[ht]
\centering
\caption{Mean and standard deviation of AUCs in test sets for 50 replications of the simulated genotype data. Model size represents the number of genetic predictors that are selected by each model. $K$ represents the number of intermediate subpopulations in the 1d linear admixture data, and the number of independent subpopulations in the independent data. $\%\Delta_{std}$ represents the relative decrease in AUC standard deviation for the predictions obtained by \texttt{pglmm}.}\label{tab:3}
\begin{tabular}{llcccccc}
  \hline
 & & \multicolumn{3}{c}{1d linear admixture} & \multicolumn{3}{c}{independent} \\ 
  \hline
  
K & Model size & \texttt{glmnetPC} & \texttt{pglmm} & $\%\Delta_{std}$ & \texttt{glmnetPC} & \texttt{pglmm} & $\%\Delta_{std}$ \\

    \hline
10 & 5 & 0.765 (0.0456) & 0.769 (0.0443) & 3.0 & 0.801 (0.0496) & 0.802 (0.0494) & 0.4 \\ 
   & 10 & 0.790 (0.0350) & 0.794 (0.0344) & 1.5 & 0.817 (0.0445) & 0.817 (0.0454) & -2.1 \\ 
   & 15 & 0.804 (0.0313) & 0.808 (0.0305) & 2.6 & 0.826 (0.0417) & 0.827 (0.0425) & -2.0 \\ 
   & 20 & 0.814 (0.0272) & 0.817 (0.0275) & -1.1 & 0.831 (0.0404) & 0.832 (0.0418) & -3.7 \\ 
   & 25 & 0.820 (0.0253) & 0.821 (0.0262) & -3.4 & 0.834 (0.0395) & 0.835 (0.0409) & -3.7 \\ 
   & 30 & 0.823 (0.0247) & 0.824 (0.0248) & -0.5 & 0.836 (0.0390) & 0.837 (0.0401) & -2.8 \\ 
   & 35 & 0.825 (0.0243) & 0.827 (0.0245) & -0.7 & 0.838 (0.0386) & 0.839 (0.0400) & -3.6 \\ 
   & 40 & 0.827 (0.0241) & 0.828 (0.0242) & -0.3 & 0.840 (0.0382) & 0.840 (0.0395) & -3.3 \\ 
   & 45 & 0.829 (0.0239) & 0.830 (0.0238) & 0.2 & 0.841 (0.0381) & 0.842 (0.0394) & -3.5 \\ 
   & 50 & 0.830 (0.0238) & 0.831 (0.0238) & -0.1 & 0.842 (0.0380) & 0.843 (0.0390) & -2.7 \\ \\ 
  20 & 5 & 0.751 (0.0431) & 0.764 (0.0419) & 2.7 & 0.771 (0.0430) & 0.807 (0.0387) & 9.9 \\ 
   & 10 & 0.775 (0.0383) & 0.788 (0.0358) & 6.5 & 0.789 (0.0387) & 0.822 (0.0355) & 8.3 \\ 
   & 15 & 0.789 (0.0356) & 0.802 (0.0313) & 12.1 & 0.801 (0.0375) & 0.830 (0.0333) & 11.0 \\ 
   & 20 & 0.798 (0.0336) & 0.811 (0.0301) & 10.4 & 0.808 (0.0368) & 0.835 (0.0316) & 14.0 \\ 
   & 25 & 0.803 (0.0327) & 0.816 (0.0299) & 8.5 & 0.815 (0.0367) & 0.838 (0.0308) & 16.1 \\ 
   & 30 & 0.807 (0.0321) & 0.819 (0.0295) & 8.0 & 0.819 (0.0361) & 0.840 (0.0305) & 15.6 \\ 
   & 35 & 0.810 (0.0315) & 0.821 (0.0297) & 5.6 & 0.822 (0.0354) & 0.842 (0.0303) & 14.4 \\ 
   & 40 & 0.812 (0.0310) & 0.823 (0.0293) & 5.5 & 0.826 (0.0346) & 0.843 (0.0301) & 13.0 \\ 
   & 45 & 0.814 (0.0309) & 0.824 (0.0293) & 5.1 & 0.829 (0.0341) & 0.844 (0.0298) & 12.4 \\ 
   & 50 & 0.816 (0.0302) & 0.825 (0.0290) & 3.9 & 0.831 (0.0336) & 0.845 (0.0297) & 11.9 \\ 
   \hline
\end{tabular}
\end{table}

\begin{table}[h]
\centering
\caption{PRS results for the UK Biobank AUC values for asthma. We report mean of AUC and standard deviation for a total of 40 random splits. For model size, we report median and interquartile range for the number of genetic predictors selected. For \texttt{pglmm}, we compare performance when model is selected using BIC or CV. For BIC, the best model is chosen based on training fit. For CV, the best model is chosen based on maximum AUC on the validation set.}\label{tab:2}


\begin{tabular}{l||l|l|l}
\hline
Model   & AUC$_{val}$   & AUC$_{test}$  & Size \\
\hline
Asthma  &  & & \\ \hline
Covariates + 10PCs & 0.5232 (0.019) & 0.5254 (0.026) & -   \\
\texttt{glmnetPC} (CV) & 0.5410 (0.018) & 0.5253 (0.027) & 67.5 (486) \\
\texttt{pglmm} (CV) & 0.5539 (0.023) & 0.5385 (0.026) & 16 (145)   \\
\texttt{pglmm} (BIC) & \multicolumn{1}{c|}{-} & 0.5452 (0.025) & 1 (1) \\
 \hline
High cholesterol &  & & \\ \hline
Covariates + 10PCs & 0.7183 (0.017) & 0.7215 (0.017) & -   \\
\texttt{glmnetPC} (CV) & 0.7196 (0.018) & 0.7196 (0.018) & 0.5 (15.5) \\
\texttt{pglmm} (CV)  & 0.7213 (0.018) & 0.7202 (0.019) & 3 (33.5)  \\
\texttt{pglmm} (BIC) & \multicolumn{1}{c|}{-} & 0.7212 (0.017) & 0 (0) \\ 
\hline
\end{tabular}
\end{table}

\clearpage
\section*{Figures}\label{sec:fig}

\end{document}


\maketitle

\clearpage
\subsection*{Appendix A. Estimation of Variance Component Parameters} \label{subsec:varcomp}
We fit the generalized linear mixed model (GLMM) \eqref{eq:model} under the assumption of no genetic association to estimate the variance components and dispersion parameters. We detail in this Appendix the AI-REML algorithm as derived by~\citet{Chen2016}.

If $\phi$ and $\bm\tau$ are known, we jointly choose $\hat{\bm{\alpha}}(\phi, \bm{\tau})$, $\hat{\bm{\gamma}}(\phi, \bm{\tau})$ and $\hat{\bm{b}}(\phi, \bm{\tau})$ to minimize  \eqref{eq:A3}, then $\hat{\bm{b}}(\phi, \bm{\tau}) = \tilde{\bm{b}}(\hat{\bm{\alpha}}(\phi, \bm{\tau}), \hat{\bm{\gamma}}(\phi, \bm{\tau}))$ because $\tilde{\bm{b}}$ maximizes $f(\bm{b})$ for given $(\bm{\alpha}, \bm{\gamma})$. Assuming that the weights in $\bm{W}$ vary slowly with the
conditional mean, the derivatives of \eqref{eq:A3} at $\bm{\gamma}=0$ with respect to $(\bm{\alpha}, \bm{b})$ are given by
\begin{align}\label{eq:A4_}
\frac{\partial ql(\bm{\alpha}, \bm{\gamma}=0, \phi, \bm{\tau} )}{\partial\bm{\alpha}} &= -\sum_{i=1}^n \frac{a_i(y_i - \mu_i)}{\phi \nu(\mu_i)}\frac{1}{g'(\mu_i)}\bm{X_}i^\intercal = -\bm{X}^\intercal\bm{W}\Delta(\bm{y}-\bm{\mu}), \nonumber \\
\frac{\partial ql(\bm{\alpha}, \bm{\gamma}=0, \phi, \bm{\tau})}{\partial\bm{b}} &= 
-\sum_{i=1}^n \frac{a_i(y_i - \mu_i)}{\phi \nu(\mu_i)}\frac{1}{g'(\mu_i)}\bm{Z_}i^\intercal + \left( \sum_{s=1}^S\tau_s\bm{V}_s \right)^{-1}\bm{b} = \left( \sum_{s=1}^S\tau_s\bm{V}_s \right)^{-1}\bm{b} - \bm{W}\Delta(\bm{y} - \bm{\mu}), \nonumber
\end{align}
where $\Delta = \textrm{diag}(g'(\mu_i))$ and $\bm{Z}_i$ is a $n\times 1$ vector of indicators such that $b_i=\bm{Z}_i\bm{b}$. Defining the working vector $\bm{\Y}$ with elements $\tilde{Y_i} = \eta_i + g'(\mu_i)(y_i - \mu_i)$, the solution of 
\begin{gather*}
\begin{cases}
\qquad \bm{X}^\intercal\bm{W}\Delta(\bm{y}-\bm{\mu}) =0 \\
\bm{W}\Delta(\bm{y} - \bm{\mu}) = \left( \sum_{s=1}^S\tau_s\bm{V}_s \right)^{-1}\bm{b}
\end{cases}
\end{gather*}
can be written as the solution to the system
\begin{align*}
\begin{bmatrix}
\bm{X^}\intercal\bm{W}\bm{X} & 
\bm{X^}\intercal\bm{W} \\
\bm{W}\bm{X} & 
\left( \sum_{s=1}^S\tau_s\bm{V}_s \right)^{-1} + \bm{W}
\end{bmatrix}
\begin{bmatrix}
\bm{\alpha} \\ 
\bm{b}
\end{bmatrix}
=
\begin{bmatrix}
\bm{X^}\intercal\bm{W}\bm{\Y}\\
\bm{W}\bm{\Y}
\end{bmatrix}.
\end{align*}
Let $\bm{\Sigma} = \bm{W}^{-1} + \sum_{s=1}^S\tau_s\bm{V}_s$, $\bm{P}=\bm{\Sigma}^{-1} - \bm{\Sigma}^{-1}\bm{X}\left(\bm{X}^\intercal\bm{\Sigma}^{-1}\bm{X}\right)^{-1}\bm{X}^\intercal\bm{\Sigma}^{-1}$, 
then 
\begin{align*}
&\begin{cases}
\hat{\bm{\alpha}} = \left(\bm{X^}\intercal\bm{\Sigma}^{-1}\bm{X}\right)^{-1}\bm{X}^{\intercal}\bm{\Sigma}^{-1} \bm{\Y} \\
\hat{\bm{b}}=\left( \sum_{s=1}^S\tau_s\bm{V}_s \right)\bm{\Sigma}^{-1}\left(\bm{\Y}-\bm{X}\hat{\bm{\alpha}}\right)
\end{cases}.
\end{align*} 
Of note, we have that
\begin{align*}
\bm{\Y}-\hat{\bm{\eta}}
&=\bm{\Y}-\bm{X}\hat{\bm{\alpha}} - \hat{\bm{b}} \nonumber \\
&= \left\{\bm{I} - \left(\sum_{s=1}^S\tau_s\bm{V}_s\right) \bm{\Sigma}^{-1} \right\}\left(\bm{\Y}-\bm{X}\hat{\bm{\alpha}} \right) \nonumber\\
&= \bm{W}^{-1}\bm{\Sigma}^{-1}\left(\bm{\Y}-\bm{X}\hat{\bm{\alpha}} \right) \nonumber\\
&= \bm{W}^{-1}\bm{P}\bm{\Y}.
\end{align*}

The log integrated quasi-likelihood function in  \eqref{eq:A3} evaluated at $(\hat{\bm{\alpha}}, \bm{\gamma}=0, \phi, \bm{\tau})$ becomes
\begin{align*}
ql(\hat{\bm{\alpha}}(\phi, \bm{\tau}), \bm{\gamma}=0, \phi, \bm{\tau}) =  &-\frac{1}{2}\text{log}\left| \sum_{s=1}^S\tau_s\bm{V}_s\bm{W} + \bm{I}\right| -\frac{1}{2}\sum_{i=1}^n\frac{a_i(y_i-\hat\mu_i)^2}{\phi\nu(\hat\mu_i)} - \frac{1}{2}\hat{\bm{b}}^\intercal\left( \sum_{s=1}^S\tau_s\bm{V}_s \right) \bm{\Sigma}^{-1}\hat{\bm{b}} \\
=&-\frac{1}{2}\text{log}\left|\bm{\Sigma}\bm{W}\right| -\frac{1}{2}(\bm{\Y} - \hat{\bm{\eta}})^\intercal\bm{W}(\bm{\Y} - \hat{\bm{\eta}}) \\
& - \frac{1}{2}\left(\bm{\Y}-\bm{X}\hat{\bm{\alpha}}\right)^\intercal\bm{\Sigma}^{-1}\left( \sum_{s=1}^S\tau_s\bm{V}_s \right) \bm{\Sigma}^{-1}\left(\bm{\Y}-\bm{X}\hat{\bm{\alpha}}\right) \\
=&-\frac{1}{2}\text{log}\left|\bm{W}\right| -\frac{1}{2}\text{log}\left|\bm{\Sigma}\right|  -\frac{1}{2}\bm{\Y}^{\intercal}\bm{P}\bm{W}^{-1}\bm{P}\bm{\Y} \\
& - \frac{1}{2}\bm{\Y}^{\intercal}\bm{P}\left( \sum_{s=1}^S\tau_s\bm{V}_s \right)\bm{P}\bm{\Y} \\
&=c -\frac{1}{2}\text{log}\left|\bm{\Sigma}\right| -\frac{1}{2}\bm{\Y}^{\intercal}\bm{P}\bm{\Sigma}\bm{P}\bm{\Y} \\
&=c -\frac{1}{2}\text{log}\left|\bm{\Sigma}\right| -\frac{1}{2}\bm{\Y}^{\intercal}\bm{P}\bm{\Y}.
\end{align*}
Similarly, the restricted maximum likelihood (REML) version is
\begin{align*}
ql_R(\hat{\bm{\alpha}}(\phi, \bm{\tau}), \bm{\gamma}=0, \phi, \bm{\tau})= 
c_R &-\frac{1}{2}\text{log}\left|\bm{\Sigma}\right| -\frac{1}{2}\text{log}\left| \bm{X}^\intercal\bm{\Sigma}^{-1}\bm{X}\right| -\frac{1}{2}\bm{\Y}^{\intercal}\bm{P}\bm{\Y}.
\end{align*}
We need to maximize ${ql_R(\hat{\bm{\alpha}}(\phi, \bm{\tau}), \bm{\gamma}=0, \phi, \bm{\tau})}$ with respect to $\phi, \bm{\tau}$. Let $\bm{V}_0 = \textrm{diag}\{a_i^{-1}\nu(\mu_i)[g'(\mu_i)^2]\} = \phi^{-1}\bm{W}^{-1}$, then $\bm{\Sigma} = \phi\bm{V}_0 + \sum_{s=1}^S \tau_s\bm{V}_s$, and the first derivatives of ${ql_R(\hat{\bm{\alpha}}(\phi, \bm{\tau}), \bm{\gamma}=0, \phi, \bm{\tau})}$ with respect to $\phi$ and $\tau_s$ are
\begin{align}
\frac{\partial ql_R(\hat{\bm{\alpha}}(\phi, \bm{\tau}), \bm{\gamma}=0, \phi, \bm{\tau})}{\partial \phi} &= \frac{1}{2}\left\{\bm{\Y}^\intercal\bm{P}\bm{V}_0\bm{P}\bm{\Y} -tr(\bm{P}\bm{V}_0) \right\} \label{eq:partialderiv1}\\
\frac{\partial ql_R(\hat{\bm{\alpha}}(\phi, \bm{\tau}), \bm{\gamma}=0, \phi, \bm{\tau})}{\partial {\tau_s}} &= \frac{1}{2}\left\{\bm{\Y}^\intercal\bm{P}\bm{V}_s\bm{P}\bm{\Y} -tr(\bm{P}\bm{V}_s) \right\}, \label{eq:partialderiv2}
\end{align}
since one can show that 
\begin{align*}
\frac{\partial\bm{P}}{\partial \phi}=-\bm{P}\bm{V}_0\bm{P}, \qquad  \frac{\partial\bm{P}}{\partial \tau_s}=-\bm{P}\bm{V}_s\bm{P}.
\end{align*}
$\hat{\phi}$ and $\hat{\bm{\tau}}$ are estimated by finding the solutions of \eqref{eq:partialderiv1} and \eqref{eq:partialderiv2} equal to zero. Let $\bm{\theta}=(\phi, \bm{\tau})$, and recall that in the REML iterative process, $\hat{\bm{\theta}}$ at the (i+1)th iteraton is updated by $\bm{\hat\theta}^{(i+1)} = \bm{\hat\theta}^{(i)} + J(\bm{\hat\theta}^{(i)})^{-1} S(\bm{\hat\theta}^{(i)})$, where $S(\bm{\theta})=\frac{\partial ql_{R}(\bm{\theta})}{\partial\bm{\theta}}$ and $J(\bm{\theta})=-\frac{\partial^2 ql_{R}(\bm\theta{})}{\partial \bm{\theta}^2}$. The elements of the observed information matrix $J(\bm{\theta})$ are 
\begin{align*}
& -\frac{\partial^2 ql_R(\hat{\bm{\alpha}}(\phi, \bm{\tau}), \bm{\gamma}=0, \phi, \bm{\tau})}{\partial \phi^2} = \bm{\Y}^\intercal\bm{P}\bm{V}_0\bm{P}\bm{V}_0\bm{P}\bm{\Y} - \frac{1}{2}tr(\bm{P}\bm{V}_0\bm{P}\bm{V}_0) \\
& -\frac{\partial^2 ql_R(\hat{\bm{\alpha}}(\phi, \bm{\tau}), \bm{\gamma}=0, \phi, \bm{\tau})}{\partial \phi \partial  \tau_s} = \bm{\Y}^\intercal\bm{P}\bm{V}_0\bm{P}\bm{V}_s\bm{P}\bm{\Y} - \frac{1}{2}tr(\bm{P}\bm{V}_0\bm{P}\bm{V}_s) \\
& -\frac{\partial^2 ql_R(\hat{\bm{\alpha}}(\phi, \bm{\tau}), \bm{\gamma}=0, \phi, \bm{\tau})}{\partial \tau_l \partial  \tau_s} = \bm{\Y}^\intercal\bm{P}\bm{V}_l\bm{P}\bm{V}_s\bm{P}\bm{\Y} - \frac{1}{2}tr(\bm{P}\bm{V}_l\bm{P}\bm{V}_s).
\end{align*}
The elements of the expected information matrix are
\begin{align*}
& E \left(-\frac{\partial^2 ql_R(\hat{\bm{\alpha}}(\phi, \bm{\tau}), \bm{\gamma}=0, \phi, \bm{\tau})}{\partial \phi^2}\right) =  \frac{1}{2}tr(\bm{P}\bm{V}_0\bm{P}\bm{V}_0)\\
& E \left(-\frac{\partial^2 ql_R(\hat{\bm{\alpha}}(\phi, \bm{\tau}), \bm{\gamma}=0, \phi, \bm{\tau})}{\partial \phi \partial  \tau_s} \right)=  \frac{1}{2}tr(\bm{P}\bm{V}_0\bm{P}\bm{V}_s)\\
& E \left(-\frac{\partial^2 ql_R(\hat{\bm{\alpha}}(\phi, \bm{\tau}), \bm{\gamma}=0, \phi, \bm{\tau})}{\partial \tau_l \partial  \tau_s} \right)= \frac{1}{2}tr(\bm{P}\bm{V}_l\bm{P}\bm{V}_s).
\end{align*}

The average information matrix $\bm{AI}$ is defined as the average of the observed information $J(\bm{\theta})$ and the expected information
\begin{align*}
\bm{AI}_{\phi\phi} &= \frac{1}{2}\bm{\Y}^\intercal\bm{P}\bm{V}_0\bm{P}\bm{V}_0\bm{P}\bm{\Y}, \\
\bm{AI}_{\phi\tau_s} &= \frac{1}{2}\bm{\Y}^\intercal\bm{P}\bm{V}_0\bm{P}\bm{V}_s\bm{P}\bm{\Y}, \\
\bm{AI}_{\tau_s\tau_l} &= \frac{1}{2}\bm{\Y}^\intercal\bm{P}\bm{V}_s\bm{P}\bm{V}_l\bm{P}\bm{\Y}.
\end{align*}

Let $\bm{\theta}$ be the variance component and dispersion parameters to estimate, that is when $\phi \ne 1$, $\bm{\theta}=(\phi, \bm{\tau}),$ and $\bm{AI}$ is a $(S + 1) \times (S+1)$ matrix. For binary data, $\phi=1$, $\bm{\theta}=\bm{\tau},$ and $\bm{AI}$ is a $S \times S$ matrix containing only $\bm{AI}_{\tau_s\tau_l}$. We use the following algorithm to estimate $\bm{\theta}$, $\bm{\alpha}$ and $\bm{b}$:

\begin{algorithm}[H]
\begin{enumerate}
    \item \textit{Initialization} \\ Fit a generalized linear model with $\bm{\tau}=0 \text{ and get } \hat{\bm{\alpha}}^{(0)} \text{ and working vector } \bm{\Y}^{(0)}$; \\
    Use $\bm{\theta}^{(0)} = Var(\bm{\Y}^{(0)})/S \text{ (if }\phi = 1\text{) or }\bm{\theta}^{(0)}=Var(\bm{\Y}^{(0)})/(S+1) \text{ (if }\phi \ne 1)$ as the initial value of $\bm{\theta}$; \\
    For each $s=0,1,...,S, \text{ update } \bm{\theta} \text{ using } \theta_s^{(1)}= \theta_s^{(0)}+2n^{-1}\{\theta_s^{(0)}\}^2(\partial ql_{R}(\bm{\theta}^{(0)})/\partial\theta_s)$; \\
    
    \item \textit{Iteration} \\
\For{$t=1,2,...,$ until convergence}{
    \qquad Update $\bm{\theta}^{(t+1)}=\bm{\theta}^{(t)} + \{\bm{AI}^{(t)}\}^{-1}(\partial ql_{R}(\bm{\theta}^{(t)})/\partial\bm{\theta})$; \\
    \qquad Calculate $\hat{\bm{\alpha}}^{(t+1)} \text{ and }\hat{\bm{b}}^{(t+1)} \text{ using } \bm{\Y}^{(t)} \text{ and } \bm{\theta}^{(t+1)}$; \\
    \qquad Update $\bm{\Y}^{(t+1)} \text{ using } \hat{\bm{\alpha}}^{(t+1)} \text{ and } \hat{\bm{b}}^{(t+1)}$;
 }
 
\end{enumerate}

 ${}$ \\ 
Convergence is defined using $2\textrm{ max}\{|\hat{\bm{\alpha}}^{(t)} - \hat{\bm{\alpha}}^{(t-1)}| / (|\hat{\bm{\alpha}}^{(t)}| + |\hat{\bm{\alpha}}^{(t-1)}|), |\hat{\bm{\theta}}^{(t)} - \hat{\bm{\theta}}^{(t-1)}|/(|\hat{\bm{\theta}}^{(t)}| + |\hat{\bm{\theta}}^{(t-1)}|)\} \le \text{tolerance}$

\caption{AI-REML algorithm}

\end{algorithm} \leavevmode\newline


\subsection*{Appendix B. Block Coordinate Descent for PQL Regularized Parameters} \label{subsec:}

Assuming that the variance components and dispersion parameters are known, we fit the full GLMM \eqref{eq:model} with lasso regularization on ${\bm{\beta}}=( {\bm{\alpha}}^\intercal , {\bm{\gamma}}^\intercal)^\intercal$ to obtain PQL regularized estimates for $\bm{\beta}$ and $\bm{b}$. At each iteration, we cycle through the coordinates and minimize the objective function \eqref{eq:objfunc} with respect to one coordinate only. Suppose we have estimates $\tilde{\bm{\beta}}$ and we wish to partially optimize \eqref{eq:objfunc} with respect to $\tilde{\bm{b}}$. The derivative at $\bm{\beta}=\tilde{\bm{\beta}}$ is given by
\begin{align}\label{eq:A4}
\frac{\partial Q_{\lambda}(\bm{\beta}, \bm{b})}{\partial\bm{b}}| _{\bm{\beta}=\tilde{\bm{\beta}}} &= 
-\sum_{i=1}^n \frac{a_i(y_i - \mu_i)}{\hat\phi \nu(\mu_i)}\frac{1}{g'(\mu_i)}\bm{Z_}i^\intercal + \left( \sum_{s=1}^S\hat\tau_s\bm{V}_s \right)^{-1}\bm{b} \nonumber \\ &
=  - {\bm{W}}\Delta(\bm{y} - \bm{\mu}) + \left( \sum_{s=1}^S\hat\tau_s\bm{V}_s \right)^{-1}\bm{b},
\end{align}
where $\Delta = \textrm{diag}(g'(\mu_i))$ and $\bm{Z}_i$ is a $n\times 1$ vector of indicators such that $b_i=\bm{Z}_i\bm{b}$. Defining the working vector $\bm{\Y}$ with elements $\tilde{Y_i} = \eta_i + g'(\mu_i)(y_i - \mu_i)$, the solution of \eqref{eq:A4} is equal to
\begin{align}\label{eq:bsolution}
\hat{\bm{b}}=\left( \sum_{s=1}^S\hat\tau_s\bm{V}_s \right){\bm{\Sigma}}^{-1}\left(\bm{\Y}-\tilde{\bm{X}}\tilde{\bm{\beta}}\right),
\end{align} 
where $\tilde{\bm{X}}=\left[\bm{X} \quad \bm{G}\right]$ and ${\bm{\Sigma}} = {\bm{W}}^{-1} + \sum_{s=1}^S\hat\tau_s\bm{V}_s$. Let $\sum_{s=1}^S\hat{\tau}_s \bm{V}_s = \bm{U}\bm{D}\bm{U}^\intercal$ be the associated eigen-spectral decomposition of the variance-covariance matrix of $\bm{b}$, where  $\bm{U}_{n\times n}$ is an orthonormal matrix of eigenvectors and $\bm{D}_{n\times n}$ is a diagonal matrix of eigenvalues, such that \eqref{eq:bsolution} can be rewritten as
\begin{align}
\label{eq:bsolution2}
\hat{\bm{b}} = \bm{U}\left(\bm{D}^{-1} + \bm{U}^\intercal\bm{W}\bm{U}\right)^{-1}\bm{U}^\intercal\bm{W}\left(\tilde{\bm{Y}} - \tilde{\bm{X}}\tilde{\bm{\beta}}\right).   
\end{align}
By rotating the random effect $\bm{\delta} = \bm{U}^\intercal\bm{b}$, we have that \eqref{eq:bsolution2} is equivalent to solving the following generalized ridge regression problem
\begin{align*}
\hat{\bm{\delta}} = \underset{\bm{\delta}}{\textrm{argmin }}\left(\tilde{\bm{Y}} - \tilde{\bm{X}}\tilde{\bm{\beta}} - \bm{U\delta}\right)^\intercal\bm{W}^{-1}\left(\tilde{\bm{Y}} - \tilde{\bm{X}}\tilde{\bm{\beta}} - \bm{U\delta}\right) + \bm{\delta}^\intercal\bm{D}^{-1}\bm{\delta}.   
\end{align*}


Consider now a coordinate descent step for $\bm{\beta}$. That is, suppose we have updates $\tilde{\bm{b}}$ and $\tilde{\bm{\beta}}_l$ for $l \ne j$, and we wish to partially optimize with respect to $\beta_j$. We would like to compute the gradient at $\beta_j = \tilde{\beta_j}$, which only exists if $\tilde\beta_j \ne 0$. If $\tilde\beta_j > 0$, then
\begin{align}\label{eq:appxB1}
\frac{\partial Q_{\lambda}(\bm{\beta}, \bm{b})}{\partial\beta_j}| _{(\bm{b}, \bm{\beta})=(\tilde{\bm{b}}, \tilde{\bm{\beta}})} &= 
-\sum_{i=1}^n \frac{a_i(y_i - \mu_i)}{\hat\phi \nu(\mu_i)}\frac{1}{g'(\mu_i)}\tilde{X}_{ij} + \lambda v_j = -\tilde{\bm{X}}_{j}^\intercal{\bm{W}}\Delta(\bm{y} - \bm{\mu}) + \lambda v_j,
\end{align}
where $\tilde{\bm{X}}_j$ is a $n\times 1$ column vector for predictor $j$.
Recall that we defined the working vector $\tilde{\bm{Y}}$ such that $\bm{y}-\bm{\mu} = {\bm{\Delta}}^{-1}(\tilde{\bm{Y}} - \tilde{\bm{X}}\tilde{\bm{\beta}} - \tilde{\bm{b}})$. Thus, plugging $\tilde{\bm{b}}$ from \eqref{eq:bsolution} and solving \eqref{eq:appxB1} leads to
\begin{gather*}
-\tilde{\bm{X}}_{j}^\intercal{\bm{W}}\left(\bm{\Y}-\tilde{\bm{X}}\tilde{\bm{\beta}} - \tilde{\bm{b}}\right) + \lambda v_j = 0 \\
\Longleftrightarrow -\tilde{\bm{X}}_{j}^\intercal{\bm{W}}\left(\bm{I} - \left(\sum_{s=1}^S\hat\tau_s\bm{V}_s\right){\bm{\Sigma}}^{-1}\right)\left(\bm{\Y}-\tilde{\bm{X}}\tilde{\bm{\beta}}\right) + \lambda v_j = 0 \\
\Longleftrightarrow -\tilde{\bm{X}}_{j}^\intercal{\bm{\Sigma}}^{-1}\left(\bm{\Y}-\tilde{\bm{X}}\tilde{\bm{\beta}}\right) + \lambda v_j = 0,
\end{gather*}
since ${\bm{\Sigma}} = \bm{W}^{-1} + \sum_{s=1}^S\hat{\tau}_s\bm{V}_s.$ Finally, isolating $\hat{\beta}_j$ yields
\begin{align}\label{eq:appxB2}
\hat{\beta}_j &= \frac{\tilde{\bm{X}_j}^\intercal{\bm{\Sigma}}^{-1}\left(\tilde{\bm{Y}}-\sum_{l\ne j}\tilde{\bm{X}}_l\tilde{\beta_l}\right) - \lambda v_j }{\tilde{\bm{X}}_j^\intercal{\bm{\Sigma}}^{-1}\tilde{\bm{X}}_j}.
\end{align}

Because the weights $\bm{W}$ are being updated at every iteration, the solution in \eqref{eq:appxB2} requires inverting a different variance-covariance matrix ${\bm{\Sigma}}$ each time we update our estimate for ${\beta_j}$. In modern large-scale genetics data sets, the sample size $n$ can be very large, thus we want to avoid costly matrices inversions. As detailed in Appendix C, we instead replace the inverse variance-covariance matrix $\bm{\Sigma}^{-1}$ in \eqref{eq:appxB2} by an upper-bound $\tilde{\bm{\Sigma}}^{-1}$~\citep{bohningMonotonicityQuadraticapproximationAlgorithms1988} and rotate the response vector $\tilde{\bm{Y}}$ and design matrix $\tilde{\bm{X}}$ by the eigenvectors of $\sum_{s=1}^S\hat{\tau}_s\bm{V}_s$, which yields
\begin{align*}
\hat{\beta}_j &= \frac{\tilde{\bm{X}_j}^\intercal\tilde{\bm{\Sigma}}^{-1}\left(\tilde{\bm{Y}}-\sum_{l\ne j}\tilde{\bm{X}}_l\tilde{\beta_l}\right) - \lambda v_j }{\tilde{\bm{X}}_j^\intercal\tilde{\bm{\Sigma}}^{-1}\tilde{\bm{X}}_j} \nonumber \\
&= \frac{\tilde{\bm{X}_j}^\intercal\left(c\bm{I}_n + \bm{UDU}^{\intercal}\right)^{-1}\left(\tilde{\bm{Y}}-\sum_{l\ne j}\tilde{\bm{X}}_l\tilde{\beta_l}\right) - \lambda v_j }{\tilde{\bm{X}}_j^\intercal\left(c\bm{I}_n + \bm{UDU}^{\intercal}\right)^{-1}\tilde{\bm{X}}_j} \nonumber \\
&= \frac{\tilde{\bm{X}_j}^\intercal\bm{U}\left(c\bm{I}_n + \bm{D}\right)^{-1}\bm{U}^{\intercal}\left(\tilde{\bm{Y}}-\sum_{l\ne j}\tilde{\bm{X}}_l\tilde{\beta_l}\right) - \lambda v_j }{\tilde{\bm{X}}_j^\intercal\bm{U}\left(c\bm{I}_n + \bm{D}\right)^{-1}\bm{U}^{\intercal}\tilde{\bm{X}}_j} \nonumber \\
&= \frac{\sum_{i=1}^n\frac{1}{c + \Lambda_i}\tilde{X}_{ij}^*\left(\tilde{Y}_i^*-\sum_{l\ne j}\tilde{X}_{il}^*\tilde{\bm{\beta}_l}\right)- \lambda v_j}{\sum_{i=1}^n\frac{1}{c + \Lambda_i}\tilde{X}_{ij}^{*^2}},
\end{align*}
where $c=4$ for binary responses, and $c=\hat\phi$ otherwise, $\Lambda_i$ are the eigenvalues of $\sum_{s=1}^S\hat{\tau}_s\bm{V}_s$, $\tilde{\bm{Y}}^* = \bm{U}^\intercal\tilde{\bm{Y}}$ and $\tilde{\bm{X}}^* = \bm{U}^\intercal\tilde{\bm{X}}$. By proceeding in a similar way for $\tilde{\beta}_j < 0$, one can show \citep{friedmanPathwiseCoordinateOptimization2007} that the coordinate-wise update for $\beta_j$ has the form
\begin{align}\label{eq:betasolution}
\hat{\beta}_j 
&= \frac{S\left(\sum_{i=1}^n\frac{1}{c + \Lambda_i}\tilde{X}_{ij}^*\left(\tilde{Y}_i^*-\sum_{l\ne j}\tilde{X}_{il}^*\tilde{\bm{\beta}_l}\right), \lambda v_j \right)}{\sum_{i=1}^n\frac{1}{c + \Lambda_i}\tilde{X}_{ij}^{*^2}},
\end{align}
where $S(z, \gamma)$ is the soft-thresholding operator:
\begin{align*}
\textrm{sign}(z)(|z| - \gamma)_{+} = 
\begin{cases}
z - \gamma & \text{if } z>0 \text{ and } \gamma < |z| \\
z + \gamma & \text{if } z<0 \text{ and } \gamma < |z| \\
0 & \text{if } \gamma \ge |z|.
\end{cases}
\end{align*}

The updates estimate for $\bm{b}$ can also be simplified by replacing the inverse variance-covariance matrix $\bm{\Sigma}^{-1}$ by $\tilde{\bm{\Sigma}}^{-1}$ in \eqref{eq:bsolution}, which yields
\begin{align*}
\hat{\bm{b}} =    \bm{U}\left(c\bm{D}^{-1} + \bm{I}_n\right)^{-1}\left(\tilde{\bm{Y}}^* - \tilde{\bm{X}}^*\tilde{\bm{\beta}}\right). 
\end{align*}

This leads to the following updates estimate for $\bm{\eta}$
\begin{align}\label{eq:eta}
\hat{\bm{\eta}}
&=\bm{\Y}- \left(\tilde{\bm{Y}} - \bm{X}\hat{\bm{\beta}} - \hat{\bm{b}}\right) \nonumber \\
&= \tilde{\bm{Y}} - \bm{U}\left\{\bm{I}_n -(c\bm{D}^{-1} + \bm{I}_n)^{-1} \right\}\left(\bm{\Y}^*-\tilde{\bm{X}}^*\hat{\bm{\beta}} \right) \nonumber\\
&= \tilde{\bm{Y}} - \bm{U}\left(c^{-1}\bm{D} + \bm{I}_n\right)^{-1}\left(\bm{\Y}^*-\tilde{\bm{X}}^*\hat{\bm{\beta}} \right).
\end{align}

The block coordinate descent algorithm to obtain regularized PQL estimates for $\bm{\beta}=(\bm{\alpha}^\intercal, \bm{\gamma}^\intercal)^\intercal$ and ${\bm{b}}$ is as follows:

\begin{algorithm}[H]
\begin{enumerate}
    \item \textit{Initialization} \\ Set $\hat{\bm{\beta}}^{(0)} = (\hat{\bm{\alpha}}^\intercal, \bm{0}^\intercal) \text{ and } \hat{\bm{b}}^{(0)} = \hat{\bm{b}}$ , where $\hat{\bm{\alpha}},\hat{\bm{b}}$ are the estimates from the AI-REML algorithm; \\
    Calculate $\hat{\bm{\eta}}^{(0)} = \tilde{\bm{X}}\hat{\bm{\beta}}^{(0)} + \hat{\bm{b}}^{0} \text{, } \tilde{\bm{Y}}^{*(0)} = \bm{U}^\intercal\bm{\Y}^{(0)} \text{ and } \tilde{\bm{X}}^{*}=\bm{U}^\intercal\tilde{\bm{X}}$;
    
    \item \textit{Iteration} \\
\For{$\lambda = \lambda_{max}$ to $\lambda_{min}$}{
\qquad \For{$t=1,2,...,$ until outer-loop convergence}{
 \qquad \For{$j=1,...,m+p$}{\qquad Calculate $$\hat{\bm{\beta}}_j^{(t)}=\frac{S\left(\sum_{i=1}^n\frac{1}{c + \Lambda_i}\tilde{X}_{ij}^*\left(\tilde{Y}_i^{*(t-1)}-\sum_{l\ne j}\tilde{X}_{il}^*\hat{\bm{\beta}_l}^{(t-1)}\right), \lambda v_j \right)}{\sum_{i=1}^n\frac{1}{c + \Lambda_i}\tilde{X}_{ij}^{*^2}},$$    \qquad until inner-loop convergence;
  }
  \qquad Calculate $\hat{\bm{\eta}}^{(t)} = \tilde{\bm{Y}}^{(t-1)} - \bm{U}\left(c^{-1}\bm{D} + \bm{I}_n\right)^{-1}\left(\bm{\Y}^{*(t-1)}-\tilde{\bm{X}}^*\hat{\bm{\beta}}^{(t)} \right)$; \\
  \qquad $\text{Update }\bm{\Y}^{(t)} \text{ and } \bm{\Y}^{*(t)} \text{ using } \hat{\bm{\eta}}^{(t)}$;
 }
    \quad Set $\hat{\bm{\beta}}^{(0)} = \hat{\bm{\beta}}^{(t)}$, $\bm{\Y}^{(0)} = \bm{\Y}^{(t)} \text{ and } \bm{\Y}^{*(0)} = \bm{\Y}^{*(t)}$ as warm starts for next $\lambda$;
 }
\end{enumerate}

 \caption{Block coordinate descent for regularized PQL estimation}
\end{algorithm} \leavevmode\newline


For inner-loop convergence, we use the same criteria as~\citet{friedmanPathwiseCoordinateOptimization2007}, that is after a complete cycle of coordinate descent we look at $$\max_j\Delta_j = \max_j\sum_{i=1}^n \frac{1}{c+\Lambda_i}\tilde{X}_{ij}^{*2}(\hat{\beta}_j^{(t-1)} - \hat{\beta}_j^{(t)})^2,$$ which measures the maximum weighted sum of squares of changes in fitted values for all coefficients. If $\max_j\Delta_j$ is smaller than tolerance, we stop the coordinate descent loop. For outer-loop convergence, we calculate the fractional change in the loss function $-l_{PQL}(\bm{\alpha}, \bm{\gamma}, \hat\phi, \hat{\bm{\tau}})$ and declare convergence if its value is smaller than tolerance.

\subsection*{Appendix C. Upper-bound for the inverse variance-covariance matrix}
We want to show that the inverse of the variance-covariance matrix $\bm{\Sigma} =  \bm{W}^{-1} + \sum_{s=1}^S\tau_s \bm{V}_s$ is upper bounded by a positive definite matrix $\tilde{\bm{\Sigma}}^{-1}$ that does not depend on the sample weights $\bm{W}$, that is, $$\bm{\Sigma}^{-1} = \left( \bm{W}^{-1} + \sum_{s=1}^S\tau_s \bm{V}_s \right)^{-1} \preceq \tilde{\bm{\Sigma}}^{-1},$$ where $\bm{W}^{-1} = \textrm{diag}\left\{ \frac{a_i}{\phi\nu(\mu_i)[g'(\mu_i)^2]}\right\}^{-1}$ and $\sum_{s=1}^S\tau_s \bm{V}_s$ are $n \times n$ positive definite matrices. 

For binary responses with logistic link and variance function equal to $\nu(\mu_i)=\mu_i(1-\mu_i)$, we have that $\bm{W}^{-1} \succeq 4 \bm{I}$. It follows that for any $h \in \mathbb{R}^n$,
\begin{align*}
h^\intercal\left( \bm{W}^{-1} + \sum_{s=1}^S\tau_s \bm{V}_s \right)h &> h^\intercal\left( 4\bm{I} + \sum_{s=1}^S\tau_s \bm{V}_s \right)h.
\end{align*}
Let $\bm{A} = \bm{W}^{-1} + \sum_{s=1}^S\tau_s \bm{V}_s$ and $\bm{B}=4\bm{I} + \sum_{s=1}^S\tau_s \bm{V}_s$ such that the previous inequality can be rewritten as $$h^\intercal \bm{A} h > h^\intercal \bm{B} h.$$ Let $\lambda^A,\lambda^B$ be the eigenvalues for $\bm{A}$ and $\bm{B}$ respectively, then 
\begin{align*}
 h^\intercal \bm{A} h \ge \lambda^A_{min} & >  \lambda_{max}^B \ge h^\intercal \bm{B} h > 0 \\
 \Longleftrightarrow \frac{1}{ \lambda_{max}^B} &> \frac{1}{ \lambda_{min}^A} \\
 \Longleftrightarrow \lambda_{min}^{B^{-1}} &> \lambda_{max}^{A^{-1}} \\
 \Longleftrightarrow h^\intercal\left( \bm{W}^{-1} + \sum_{s=1}^S\tau_s \bm{V}_s \right)^{-1}h &< h^\intercal\left( 4\bm{I} + \sum_{s=1}^S\tau_s \bm{V}_s \right)^{-1}h.
\end{align*}
Thus, we have $$\bm{\Sigma}^{-1} = \left( \bm{W}^{-1} + \sum_{s=1}^S\tau_s \bm{V}_s \right)^{-1} \preceq \left( 4\bm{I} + \sum_{s=1}^S\tau_s \bm{V}_s \right)^{-1} \equiv \tilde{\bm{\Sigma}}^{-1}.$$

\bibliographystyle{apalike}
\bibliography{biblio}